\RequirePackage{iftex}
\ifPDFTeX
  \pdfoutput=1
\fi
\documentclass[11pt]{article}

\usepackage[preprint]{acl}
\usepackage{times}
\usepackage{latexsym}
\usepackage{inconsolata}
\ifPDFTeX
\usepackage[utf8]{inputenc}
\fi
\usepackage[T1]{fontenc}
\hypersetup{
  hidelinks,
  pdftitle={CMASS: A ClimbMix Agentic Search Suite for Evaluating Search Agents Across Heterogeneous Tasks},
  pdfauthor={Sahel Sharifymoghaddam, Lingwei Gu, Yijun Ge, Jimmy Lin}
}
\usepackage{url}
\usepackage{booktabs}
\usepackage{amsfonts}
\usepackage{nicefrac}
\usepackage{microtype}
\usepackage{xcolor}
\usepackage{xspace}
\usepackage{enumitem}
\usepackage{cleveref}
\usepackage{subcaption}
\usepackage{graphicx}
\usepackage{mdframed}
\usepackage{multirow}
\usepackage{placeins}

\usepackage{tcolorbox}
\usepackage{tabularx}
\newcommand{\tablestyle}{%
  \fontsize{9pt}{11pt}\selectfont
  \setlength{\tabcolsep}{4pt}%
  \renewcommand{\arraystretch}{1}%
}
\tcbuselibrary{skins}

\definecolor{hopreq}{HTML}{D9EAD3}
\definecolor{hopconf}{HTML}{CFE2F3}
\definecolor{hopung}{HTML}{F4CCCC}
\definecolor{hoppart}{HTML}{FFF2CC}

\newtcbox{\ungbox}{on line, boxrule=0pt, boxsep=0pt,
left=3pt, right=3pt, top=1.5pt, bottom=1.5pt, arc=3pt, colback=hopung, colframe=hopung}
\newtcbox{\partbox}{on line, boxrule=0pt, boxsep=0pt,
left=3pt, right=3pt, top=1.5pt, bottom=1.5pt, arc=3pt, colback=hoppart, colframe=hoppart}

\newtcbox{\reqbox}{on line, boxrule=0pt, boxsep=0pt,
left=3pt, right=3pt, top=1.5pt, bottom=1.5pt, arc=3pt, colback=hopreq, colframe=hopreq}
\newtcbox{\confbox}{on line, boxrule=0pt, boxsep=0pt,
left=3pt, right=3pt, top=1.5pt, bottom=1.5pt, arc=3pt, colback=hopconf, colframe=hopconf}

\newcommand{\docid}[1]{{\ttfamily #1}}

\newcolumntype{T}{>{\hsize=1.45\hsize\raggedright\arraybackslash}X}
\newcolumntype{D}{>{\raggedright\arraybackslash}p{75pt}}

\newcommand{\projectlink}{\url{https://cmass.tech}}
\newcommand{\ignore}[1]{}

\newcommand{\bcp}[0]{BCP\xspace}

\newcommand{\cm}[0]{ClimbMix\xspace}
\newcommand{\cmass}{CMASS\xspace}

\newcommand{\piika}[0]{\textsc{Piika}\xspace}

\title{CMASS: A ClimbMix Agentic Search Suite for Evaluating Search Agents Across Heterogeneous Tasks}

\author{Sahel Sharifymoghaddam\thanks{Equal contribution.} \quad Lingwei Gu\footnotemark[1] \quad Yijun Ge \quad Jimmy Lin\\
David R. Cheriton School of Computer Science\\
University of Waterloo}

\usepackage{tikz}
\usetikzlibrary{positioning,arrows.meta,calc,fit,backgrounds}
\newcommand{\hotpot}[0]{HotpotQA\xspace}
\newcommand{\trqa}[0]{TRQA\xspace}
\newcommand{\dsqa}[0]{DSQA\xspace}

\newcolumntype{L}{>{\raggedright\arraybackslash}X}

\definecolor{stPrep}{HTML}{EDE7F6}
\definecolor{stDecomp}{HTML}{E3F2FD}
\definecolor{stGround}{HTML}{E8F5E9}
\definecolor{stCheck}{HTML}{FFF8E1}
\definecolor{stRepair}{HTML}{FCE4EC}
\definecolor{stVerify}{HTML}{E0F7FA}
\definecolor{stHuman}{HTML}{FFF3E0}
\definecolor{stRelease}{HTML}{F1F8E9}
\definecolor{stSide}{HTML}{F5F5F5}
\definecolor{stEdge}{HTML}{607D8B}

\begin{document}

\raggedbottom
\widowpenalty=10000
\maketitle

\begin{abstract}
Evaluating advances in agentic search requires testing whether methods generalize across tasks and corpus scale.
Changing web search services impede controlled comparisons, while small corpora assembled around benchmark questions provide a limited test of generalization.
We introduce \cmass, the ClimbMix Agentic Search Suite, which projects three search benchmarks---BrowseComp-Plus, Total Recall QA, and DeepSearchQA---and the multi-hop question-answering dataset HotpotQA onto a shared, fixed corpus.
The target, \cm, is a pre-training corpus of 553M documents and approximately 400B tokens, assembled independently of the benchmark questions.
Our projection pipeline verifies grounding, answer uniqueness, and set completeness through independent agent audits, controlled revision, and human review.
The 423 questions span entity identification, multi-hop reasoning, exhaustive answer sets, and aggregation.
Evaluations of three models expose substantial headroom for open-weight agents and show that high answer accuracy can coexist with incomplete evidence coverage.
By enabling measurement of accuracy alongside evidence coverage, search effort, and closed-book performance, \cmass{} supports developing agents that find sufficient evidence and produce complete answers efficiently.
The pipeline code and dataset are available at \projectlink{}.
\end{abstract}

\section{Introduction}
\label{sec:intro}

Agentic search systems must retrieve evidence and reason over it to answer complex questions.
Improving these systems requires understanding which retrieval interfaces and search strategies help agents find and use the evidence they need.
Live web search complicates this evaluation: pages and rankings change, while proprietary search services hide how evidence is retrieved.
An agent's score can therefore change because of the search service or the available evidence, even when the agent itself is unchanged.

BrowseComp-Plus (BCP; \citealp{chen2026browsecompplus}) has opened new avenues for controlled evaluation of these systems by replacing live web search with a fixed corpus and providing query--document relevance judgments.
Researchers can vary agents and retrieval methods independently while holding the evidence source constant, and measure retrieval effectiveness alongside answer quality.
On BCP, Pi-Serini showed that well-configured BM25 with sufficient retrieval depth can power strong agent effectiveness~\citep{hsu2026piserini}, while direct corpus interaction (DCI) showcased the promise of exploring corpus text through shell tools~\citep{li2026dci}.
BCP has also supported research on training agents through reinforcement learning~\citep{sun2025contextfolding,chang2026karl}.

Together, these advances bring an important question into focus: do alternative search interfaces and different agent training methods yield consistent gains across tasks and corpus sizes?
BCP's 100K-document corpus combines benchmark-specific evidence with hard negatives, and evaluation on BCP alone does not test transfer to exhaustive set enumeration or aggregation.
DCI and subsequent work already reveal scaling challenges as the search space expands~\citep{li2026dci,zhuang2026rise}.
The natural next step is to test these approaches across a broader range of tasks on substantially larger collections assembled independently of benchmark questions.

We introduce \cmass, the \emph{ClimbMix Agentic Search Suite}, to take that step.
\cmass{} projects \bcp{}, Total Recall QA (TRQA; \citealp{rafiee2026trqa}), DeepSearchQA (DSQA; \citealp{gupta2026deepsearchqa}), and HotpotQA~\citep{yang2018hotpotqa} onto \cm{}~\citep{DiaoShizhe_etal_NeurIPS2025}, a pre-training corpus of approximately 400B tokens assembled independently of the questions.
With 553M documents---roughly $5{,}500\times$ the size of BCP's 100K-document corpus---\cmass{} brings controlled agentic search evaluation to a dramatically larger scale across entity identification, multi-hop reasoning, exhaustive set enumeration, and aggregation.

\cmass{} builds on our initial BCP projection onto \cm{}, expanding to four datasets and strengthening the pipeline with controlled question revision, adversarial checks of answer uniqueness and set completeness, and more comprehensive query--document relevance judgments (qrels).
The pipeline combines these additions with evidence grounding and human review.
The resulting questions have verified answers and qrels, with held-out hop-level annotations supporting finer-grained evaluation of evidence coverage.
As \cmass{} expands, Appendix~\ref{app:version-history} records dated changes to the paper and released artifacts.

Our initial evaluations point to clear opportunities for advancing agentic search.
They reveal substantial headroom for open-weight agents and show that high answer accuracy can coexist with incomplete evidence coverage.
Supplying the supporting documents directly resolves 32 of the strongest model's 34 retrieval-enabled failures, identifying evidence acquisition and presentation as promising targets for improvement.
\cmass{} enables researchers to measure progress on these targets across tasks, with closed-book controls and explicit measures of answer completeness and search effort.
Our main contributions are:

\begin{itemize}
\item \textbf{\cmass{} v1.0.}
A suite of four datasets projected onto \cm{}, containing 423 verified question--answer (QA) pairs and 40{,}736 qrels.
Section~\ref{sec:released-artifacts} summarizes the release.
\item \textbf{Diagnostic evaluation.}
Evaluations reveal substantial headroom for open-weight agents and show that high answer accuracy can coexist with incomplete evidence coverage.
Document recall alone does not capture hop coverage or answer accuracy.
Supplying evidence enables correct answers to missed questions, highlighting retrieval quality and evidence presentation as priorities for future systems.
\item \textbf{Reusable projection pipeline.}
A dataset- and corpus-agnostic projection pipeline, with analyses of source defects, grounding failures, and controlled revision during suite construction.
\end{itemize}

\section{Design Rationale}
\label{sec:rationale}

Below, we explain the rationale for three design choices: projecting existing questions, using one independently assembled corpus, and grounding every constraint needed to justify the answer.

\paragraph{Starting from existing benchmarks.}
\bcp{} and \hotpot{} are established benchmarks widely used in research.
The more recent \trqa{} and \dsqa{} broaden task coverage with aggregation and exhaustive answer sets.
We chose projection over constructing a new benchmark from scratch to preserve this task diversity and build on existing benchmark design within a shared retrieval environment.

\paragraph{Target corpus.}
We selected NVIDIA's \cm~\citep{DiaoShizhe_etal_NeurIPS2025}, whose construction balances data quality with broad coverage of topics and domains.
The CLIMB framework uses semantic clustering and filtering to form 20 document groups, then learns a mixture for pre-training from scratch.
Appendix~\ref{app:dups} describes the corpus's source collections and construction.
Two properties make it suitable for this evaluation:

\begin{itemize}

\item Its 553M documents provide a large search space, while the 559\,GB Lucene index remains practical for academic research.
This permits retrieval experiments over hundreds of millions of documents without requiring access to a commercial search service.

\item Its use in pre-training (nanochat; \citealp{nanochat}), QA projection (NanoKnow; \citealp{gu2026nanoknow}), and retrieval evaluation (TREC RAG 2026\footnote{\url{https://trec-rag.github.io}}) enables studies of parametric and retrieved knowledge over the same information source.
Researchers can compare what a model answers from its parameters with what it answers using search while controlling the corpus from which both forms of knowledge are drawn.
\looseness=-1

\end{itemize}

\paragraph{Projecting by reasoning chains.}
Finding a passage containing the supplied answer does not establish that the whole question is supported.
We link question constraints to reasoning hops so that an audit can identify the particular premise that lacks support.
\emph{Inclusion} establishes why a candidate satisfies a constraint; \emph{exclusion} establishes why an alternative fails one.
Every factual constraint in the accepted wording must be grounded, including confirmatory facts that do not further narrow the answer.
For aggregates, this also requires evidence for the qualifying members and values used in the calculation.
These checks apply across datasets, with dataset-specific revision limits.

\begin{figure*}[tp]
\centering
\includegraphics[width=\textwidth]{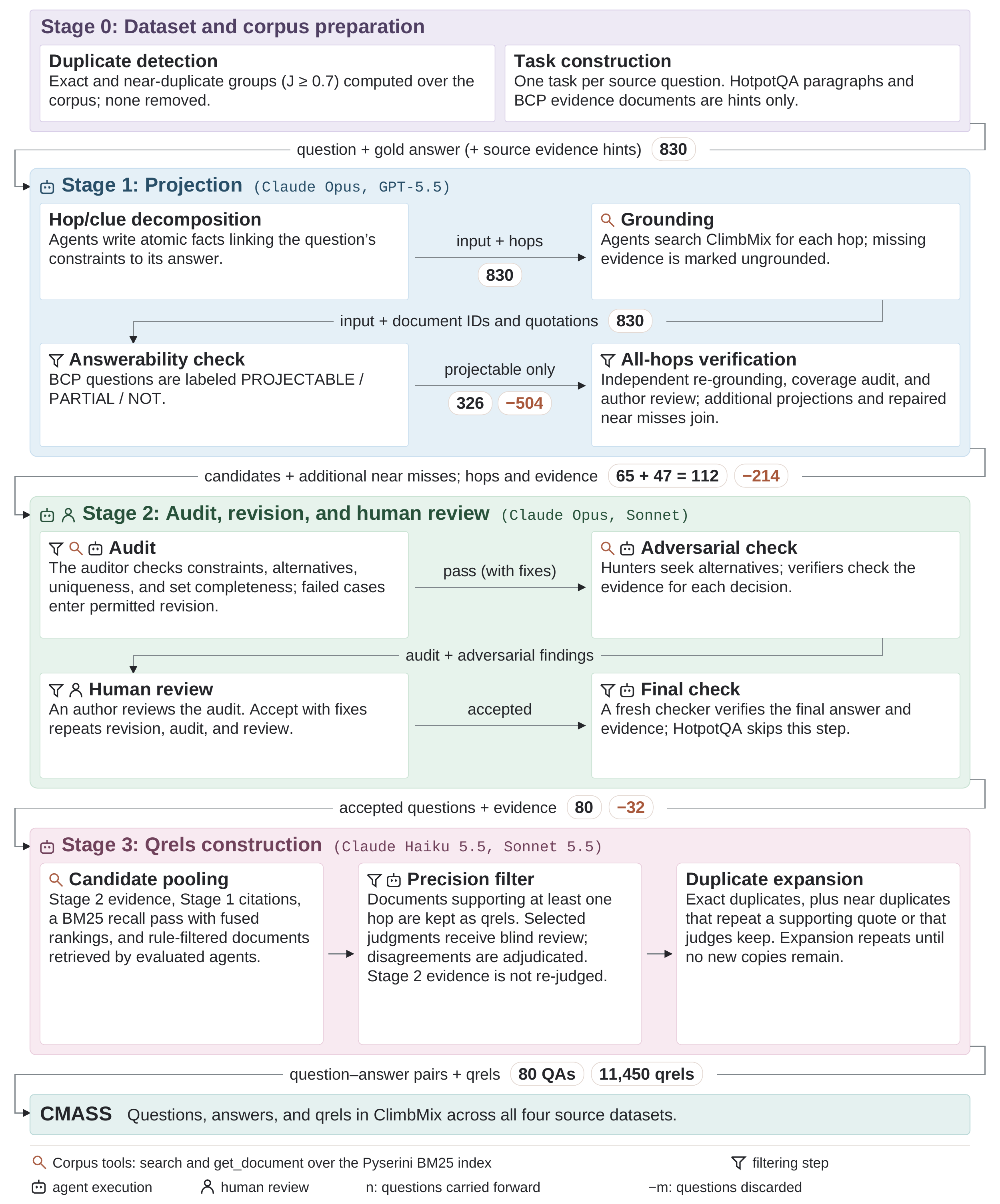}
\caption{The CMASS projection pipeline, with counts shown for BCP.
Magnifier icons indicate use of \texttt{search} and \texttt{get\_document} over the Pyserini BM25 index; funnel, robot, and person icons denote filtering, agent execution, and human review.
Count badges show questions carried forward ($n$) or discarded ($-m$).
Figure~\ref{revised:fig:stage3} details Stage~2's audit, revision, and human verification loop.}
\label{fig:pipeline-overview}
\end{figure*}

\section{Methodology}
\label{section:methodology}
The dataset- and corpus-agnostic pipeline takes question--answer pairs and a retrieval endpoint supporting search and full-document access.
Source qrels provide optional evidence hints, and the endpoint can use BM25 or another retriever.

For \cmass{}, a Pyserini~\citep{lins2021pyserini} BM25 server exposes \texttt{search} for ranked results and \texttt{get\_document} for full text by ID.
Projection and evaluation agents share the corpus and retrieval service, keeping the available evidence fixed.

Below, we detail the four pipeline stages summarized in \Cref{fig:pipeline-overview}.
Stage~0 prepares corpus and task records; Stage~1 decomposes and grounds questions; Stage~2 audits and revises them with human review; and Stage~3 constructs qrels from accepted evidence, a recall pass, and documents retrieved by evaluated agents.
Each stage preserves per-question evidence and decisions for inspection and resumption.

\paragraph{Stage 0: Dataset and corpus preparation.}
We prepare corpus metadata and question records in two steps.
First, we characterize document lengths and identify exact and near-duplicate groups within the target corpus (\emph{duplicate detection}).
Exact duplicates match after trimming leading and trailing whitespace; near-duplicate candidates are detected with MinHash and checked using Jaccard similarity over 5-gram shingles at a threshold of $0.7$.
No documents are removed: Stage~3 uses these groups to credit retrieved documents that provide equivalent evidence.
Appendix~\ref{app:dups} provides the corpus analysis and implementation details.

\begingroup
\widowpenalty=10000
Next, we create one task per source question from \bcp{}, \dsqa{}, \trqa{}, and a subset of \hotpot{} bridge questions (\emph{task construction}).
Each task contains the source question and gold answer.
\hotpot{} tasks include the source supporting paragraphs and \bcp{} tasks the source's evidence documents; both help recover the intended reasoning chain but serve as hints rather than grounding evidence.
All accepted evidence must come from~\cm{}.
\par\endgroup

\begin{figure*}[t]
\centering
\includegraphics[width=\textwidth]{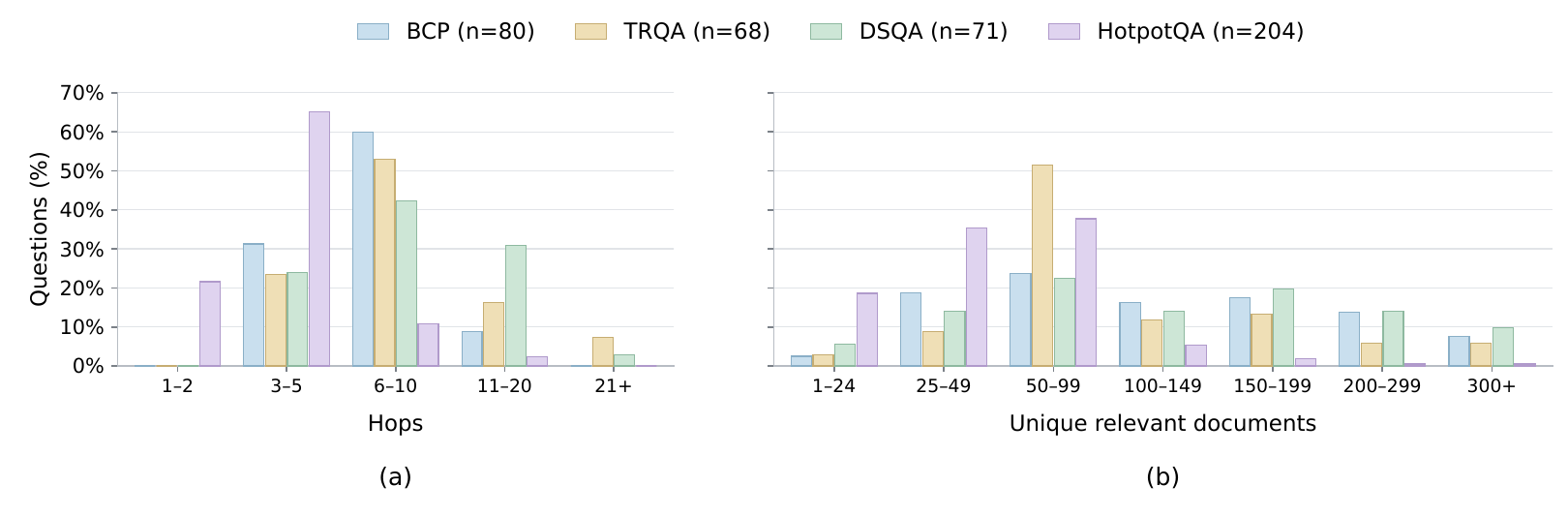}
\caption{Distributions of (a) total retained hops and (b) unique relevant documents per question in the final projections.
Each dataset is normalized by its number of final questions: BCP (80), TRQA (68), DSQA (71), and HotpotQA (204).
Hop counts include all roles; each relevant document is counted once per question.}
\label{fig:projection-stats}
\end{figure*}

\paragraph{Stage 1: Projection.}
Agents decompose each task into atomic facts, or \emph{hops}, connecting its constraints to its answer (\emph{hop/clue decomposition}).
They search \cm{} with BM25, reformulate queries, and read retrieved documents to ground each factual hop, recording the supporting document identifiers or marking the evidence as missing (\emph{grounding}); computational hops record an operation over grounded inputs.

They then judge whether the supplied answer can be derived from the grounded hops (\emph{answerability check}).
For BCP, \textsc{Projectable} means the grounded facts uniquely determine the answer; \textsc{Partial} means they leave ambiguity or weak support for a decisive step; and \textsc{Not} means a decisive fact is missing or the corpus contradicts the answer.
Additional dataset-specific verification steps then select candidates (\emph{all-hops verification}).
Quoted passages are later checked against their documents and read in context, since matching text alone does not establish the requested entity, relation, or temporal qualifier.

Passing questions form the projected candidate list; near misses are retained separately for audit and possible revision.
Both groups enter Stage~2 with their hops and evidence.
Figure~\ref{fig:projection-stats}(a) shows the final number of retained hops per question for each dataset.

The answerability criteria depend on the answer type.
For \bcp{} questions, the agent requires a supported chain to the supplied answer; for \dsqa{} sets, at least one gold member must satisfy every constraint on grounded evidence.
For \trqa{} aggregates, the full set of members and their values must support the numerical result.
The DSQA screen is preliminary: full set completeness is checked in Stage~2.
Most projections use Claude Opus; 11 additional BCP candidates are projected independently with GPT-5.5 and audited by Claude Opus.

\paragraph{Stage 2: Audit, revision, and human review.}
An \emph{auditor agent} checks constraints, searches for plausible alternatives, and assesses answer accuracy, uniqueness, and set completeness (\emph{audit}).
The auditor types each hop as \emph{required} if removing it makes the answer ambiguous or underdetermined, and \emph{confirmatory} if it corroborates a constraint without further narrowing the answer.
Required hops establish inclusion or exclusion.
Table~\ref{tab:hop-role-rates} summarizes these roles and their qrels.

\begin{table}[!b]
\tablestyle
\centering
\begin{tabular}{@{}llrr@{}}
\toprule
Dataset & Role & Hops (\%) & Qrels (\%) \\
\midrule
TRQA & Inclusion & 65.0 & 68.9 \\
 & Exclusion & 35.0 & 31.1 \\
\midrule
DSQA & Inclusion & 64.8 & 72.7 \\
 & Exclusion & 35.2 & 27.3 \\
\midrule
BCP & Required & 70.4 & 76.5 \\
 & Confirmatory & 29.6 & 23.5 \\
\bottomrule
\end{tabular}
\caption{Hop and qrel percentages pooled over 48 TRQA aggregation, 39 DSQA set-answer, and 80 BCP questions.
Documents count once per question.}
\label{tab:hop-role-rates}
\end{table}

The auditor returns \emph{pass} when the candidate meets the audit criteria without changes, \emph{pass with fixes} when adjusting hops or grounding suffices without changing the QA, and \emph{fail} otherwise.
For each plausible alternative, it must establish whether the candidate satisfies the constraints or fails at least one on grounded evidence.
An unresolved alternative candidate can undermine a single answer's uniqueness or indicate a missing set member.
Retrieval failure supports neither inclusion nor exclusion, so such alternatives block acceptance.

For sets and aggregates, completeness can rest on a corpus document listing all members or giving the set size, or on an explicitly fixed domain whose members are all checked.
Otherwise, the audit searches through list documents, named candidates, and category descriptions, requiring that no plausible candidate remain unresolved.
This search-based acceptance is a best-effort judgment, not proof that every document was checked.

An independent \emph{hunter agent} searches passing questions for alternative candidates and evidence for their inclusion or exclusion.
A separate \emph{verifier agent} checks whether the cited evidence supports each decision (\emph{adversarial check}).
One author reviews the audit and adversarial findings (\emph{human review}; Appendix~\ref{app:review-ui}).
The reviewer inspects the question, gold answer, constraint-to-hop mapping, quotations, and candidate decisions, with revised wording shown beside its source version.
Human acceptance applies to the reviewed wording and evidence; it does not automatically transfer to a later revision.
\looseness=-1

An audit failure or a human verdict of \emph{accept with fixes} triggers another retrieval attempt for missing evidence before any permitted QA revision.
Revised QAs repeat the audit, adversarial checks, and human review; rejected or unsuccessfully repaired QAs are excluded.
A verified adversarial refutation permits one more repair attempt; a second excludes the question.

After human acceptance, \bcp{}, \trqa{}, and \dsqa{} undergo an additional agent check because their revision rules can remove constraints or change answers (\emph{final check}).
\hotpot{} skips this post-review check.
If the checker finds a repairable gap in the answer or evidence, the QA returns to revision and review; unresolved failures lead to exclusion.

Revision rules depend on the dataset.
\hotpot{} permits wording corrections or one grounded qualifier while preserving the answer.
For \bcp{}, Stage~1 retained questions receive evidence-only repairs; near misses may receive a clue correction or removal, a range adjustment, or one grounded qualifier.
Their answers are preserved except for typo corrections.
\trqa{} permits a grounded qualifier; when the original question's wording makes the answer trivial, the requested attribute may also change.
\dsqa{} permits source-clause removal, grounded qualifiers, and year or threshold adjustments.
Recomputing set or aggregate answers still requires the full grounding and \mbox{completeness checks}.

Appendix~\ref{app:revised-pipeline} details the audit, failure reasons, and dataset-specific revision limits (Table~\ref{revised:tab:adapters}).

\paragraph{Stage 3: Qrels construction.}
For each verified question, qrels are the union of documents supporting its reasoning hops.
For computational hops, documents ground the factual inputs; the operation itself is verified from those inputs.

Candidates come from four sources (\emph{candidate pooling}).
Documents accepted as evidence in Stage~2 are kept without further judgment.
Documents cited in Stage~1 and, for \bcp{}, the qrels of an earlier release are judged like any other candidate.
A recall pass adds new candidates: an agent writes several queries per hop, and their BM25 rankings are fused with reciprocal rank fusion~\citep{cormack2009rrf} and filtered by hop-specific anchor terms.

Finally, following the pooling practice of TREC~\citep{sparckjones1975ideal,voorhees2005trec}, we add the documents that the three agents evaluated in \Cref{sec:evaluation} surfaced or opened.
To keep this tractable, a pooled document is judged only if its text matches one of the question's hop rules, which are literal phrase patterns written for each hop and required to match every document already judged relevant to it.

Judge agents read each candidate in full and decide which hops of the question it supports, against hop standards written before any candidate is read (\emph{precision filter}).
A document remains relevant if it supports at least one hop.
Judging is tiered: Claude Haiku~5.5 is the first judge for candidates requiring judgment, and Claude Sonnet~5.5 writes the hop standards and recall queries, re-judges selected candidates without seeing the first verdict, and adjudicates disagreements.

We then add exact duplicates of each relevant document, which inherit its hops, and near duplicates with $5$-gram Jaccard similarity of at least $0.7$ (\emph{duplicate expansion}).
A near duplicate that repeats a verbatim supporting quotation of its original inherits that hop without a separate judgment.
Other near duplicates require agent verification before inclusion.
Expansion repeats over the added documents until no new duplicate appears.
Appendix~\ref{app:stage3-details} describes the judging process and system pool and reports the source of every qrel.

In Tables~\ref{tab:hop-role-rates} and~\ref{revised:tab:yield}, a qrel is exclusion-only if every hop it supports is an exclusion hop; otherwise, it counts as inclusion.
For BCP, a qrel is confirmatory-only if every supported hop is confirmatory, and required if it supports any required hop.
Percentages pool hops and question--document pairs rather than averaging per-question rates.

\section{Released Artifacts}
\label{sec:released-artifacts}
\cmass{} v1.0 contains 423 verified English question--answer pairs and 40{,}736 query--document relevance judgments over the shared 553M-document \cm{} corpus.
The four datasets cover entity identification, multi-hop reasoning, exhaustive answer sets, and aggregation; all questions are used for evaluation.
Table~\ref{revised:tab:yield} summarizes the question counts, qrels, and revision categories.
Each qrel is a distinct question--document pair, including accepted exact and near duplicates.
Each QA belongs to one revision category, with answer changes taking precedence over question edits.
Figure~\ref{fig:projection-stats} shows the distributions of hops and relevant documents per question.

\begin{table}[!t]
\tablestyle
\setlength{\tabcolsep}{1.75pt}
\centering
\begin{tabular}{@{}lrrrrrrrr@{}}
\toprule
\multirow{2}{*}{Dataset} & \multicolumn{6}{c}{QA} & \multicolumn{2}{c}{Qrels} \\
\cmidrule(lr){2-7}\cmidrule(l){8-9}
& Total & Un. & Re. & C.R. & Na. & A.C. & In. & Ex. \\
\midrule
HotpotQA & 204 & 144 & 13 & 0 & 47 & 0 & 10{,}195 & 895 \\
TRQA & 68 & 31 & 5 & 0 & 16 & 16 & 5{,}555 & 2{,}752 \\
DSQA & 71 & 31 & 9 & 13 & 12 & 6 & 6{,}979 & 2{,}910 \\
BCP & 80 & 37 & 15 & 10 & 17 & 1 & 10{,}165 & 1{,}285 \\
\midrule
Total & 423 & 243 & 42 & 23 & 92 & 23 & 32{,}894 & 7{,}842 \\
\bottomrule
\end{tabular}
\caption{CMASS v1.0 release statistics and revision categories.
Un.: Unchanged; Re.: Reworded; C.R.: Constraint removed; Na.: Narrowed; A.C.: Answer changed; In.: Inclusion; Ex.: Exclusion.
}
\label{revised:tab:yield}
\end{table}

The release provides final questions, reference answers, question-level qrels, and inclusion/exclusion annotations; DSQA also includes structured answer sets with accepted aliases.
The projection pipeline code and dataset are available at \projectlink{}.
Document identifiers refer to the original \cm{} release, whose text is distributed separately.
We withhold hop decompositions and per-hop qrels to limit tuning to intermediate reasoning and will score submitted runs using these annotations (Appendix~\ref{app:evaluation-prompts}).

\section{\cmass{} Evaluation Results}
\label{sec:evaluation}

\cmass{} exposes a striking gap between answering correctly and retrieving the evidence needed to support the answer.
Our evaluation reveals substantial headroom for open-weight agents and concrete opportunities to improve even the strongest model.
We evaluate all 423 questions with GPT-6.1 Sol at high reasoning effort, Gemma~4 31B IT, and Qwen~3.5 9B, using closed-book answering without tools and retrieval-enabled answering with \piika{}, a search-agent framework, over \cm{}.
\piika{} follows the agent design of Pi-Serini~\citep{hsu2026piserini}, using BM25 search and document reads through the corpus endpoint described in \Cref{section:methodology}.
We report one run per model, dataset, and condition; comparisons are descriptive, without estimates of variability across repeated runs.
Appendix~\ref{app:evaluation-prompts} provides the answering and judging prompts.
\looseness=-1

\begin{table}[!t]
\tablestyle
\setlength{\tabcolsep}{1.7pt}
\centering
\begin{tabular}{@{}llrrrrrr@{}}
\toprule
& & \multicolumn{5}{c}{Retrieval tools} & \multicolumn{1}{r}{\makebox[0pt][r]{\shortstack[r]{Closed-\\book}}} \\
\cmidrule(lr){3-7}\cmidrule(l){8-8}
& Model & R@all & P@all & H.~G. & T.~C. & Acc. & Acc. \\
\midrule
\multicolumn{8}{c}{\textit{BCP (80 questions)}} \\
(a1) & GPT-6.1 Sol & 12.1 & 38.2 & 75.5 & 21.5 & 85.0 & 52.5 \\
(a2) & Gemma~4 31B IT & 4.2 & 10.9 & 31.4 & 23.5 & 27.5 & 16.3 \\
(a3) & Qwen~3.5 9B & 4.4 & 9.2 & 31.4 & 49.9 & 16.3 & 5.0 \\
\midrule
\multicolumn{8}{c}{\textit{TRQA (68 questions)}} \\
(b1) & GPT-6.1 Sol & 10.6 & 70.9 & 79.7 & 8.6 & 94.1 & 86.8 \\
(b2) & Gemma~4 31B IT & 17.0 & 59.2 & 82.7 & 12.4 & 66.2 & 50.0 \\
(b3) & Qwen~3.5 9B & 18.6 & 59.4 & 85.8 & 19.7 & 57.4 & 39.7 \\
\midrule
\multicolumn{8}{c}{\textit{DSQA (71 questions)}} \\
(c1) & GPT-6.1 Sol & 15.0 & 50.5 & 75.7 & 17.9 & 90.1 & 70.4 \\
(c2) & Gemma~4 31B IT & 17.4 & 33.2 & 68.1 & 21.5 & 38.0 & 23.9 \\
(c3) & Qwen~3.5 9B & 19.8 & 35.3 & 72.0 & 37.1 & 39.4 & 12.7 \\
\midrule
\multicolumn{8}{c}{\textit{HotpotQA (204 questions)}} \\
(d1) & GPT-6.1 Sol & 17.2 & 57.9 & 90.2 & 6.7 & 94.6 & 89.7 \\
(d2) & Gemma~4 31B IT & 18.1 & 44.4 & 78.4 & 10.2 & 65.7 & 36.8 \\
(d3) & Qwen~3.5 9B & 19.7 & 44.4 & 83.9 & 17.0 & 70.1 & 27.5 \\
\bottomrule
\end{tabular}
\caption{Retrieval-tool and closed-book results on \cmass{}.
R@all: Recall@all; P@all: Precision@all; H.~G.: hop groundedness; Acc.: accuracy (all \%).
T.~C.: mean search and document-read tool calls per question.
Sol uses high reasoning effort.
DSQA uses strict exact-set equality.}
\label{tab:cmass-evaluation}
\end{table}

In our evaluations, we measure answer accuracy, Recall@all, Precision@all, hop groundedness, and tool calls.
Accuracy is the percentage of all questions answered correctly, with unanswered questions counted as failures.
DSQA uses strict exact-set equality, requiring every gold member and no incorrect additions, regardless of order.
The 32 single-answer DSQA questions are scored as singleton sets; Table~\ref{tab:hop-role-rates} analyzes the other 39 set-answer questions.
All 68 TRQA questions, including the 20 outside that table's aggregation subset, use the semantic-equivalence judge in Figure~\ref{fig:judge-prompt}.

Recall@all measures the percentage of a question's qrel documents returned in search results, counting each document once across calls.
Precision@all measures the percentage of these distinct retrieved documents that are in the question's qrels.
Unjudged documents count as non-relevant, so incomplete qrels can underestimate precision and either overestimate or underestimate recall.

Hop groundedness uses our withheld hop-level qrels to measure the percentage of each question's hops supported by at least one retrieved document.
Search effort counts tool calls (search and document reads).
Table~\ref{tab:cmass-evaluation} reports all five metrics averaged over all questions in each dataset.
Recall, precision, and hop groundedness use the final qrels, including judged evidence pooled from the three evaluated agents.
Four findings show where the next gains in agentic search can come from.

\paragraph{Open-weight agents have substantial room to grow.}
On the BCP subset, Gemma and Qwen achieve 27.5\% and 16.3\% accuracy, versus Sol's 85.0\%, while making more tool calls (a1--a3).
The evidence annotations help locate this gap: their hop groundedness is 31.4\% for both models, compared with 75.5\% for Sol.
These comparisons do not isolate search policies from reasoning ability or prior knowledge.

Answer completeness presents another challenge: accepting incomplete but otherwise correct DSQA sets raises Gemma's and Qwen's accuracy by 15.5 and 14.1 percentage points, respectively, compared with 2.8 points for Sol (Appendix~\ref{app:supplementary-results}).
These open-weight models are more competitive when applications accept a correct subset; Sol's advantage is larger when all qualifying members must be identified, as required in high-stakes applications such as exhaustive reviews of legal cases.
\looseness=-1

\paragraph{High accuracy can conceal incomplete search.}
Even Sol's strong answer scores leave room for better search: it achieves 90.1\% accuracy on DSQA while covering only 75.7\% of hops on average (c1).
Closed-book controls reveal how much agents can answer without searching: on HotpotQA, Sol achieves 89.7\% closed-book accuracy, compared with 94.6\% with retrieval (d1).
This ability extends to revised questions: its closed-book accuracy exceeds the share of unchanged QAs on every dataset (Table~\ref{revised:tab:yield}).
However, these closed-book results do not distinguish memorized questions from factual knowledge and inference.

\paragraph{Document recall tells only part of the story.}
On HotpotQA, Sol retrieves 17.2\% of qrel documents yet grounds 90.2\% of hops (d1): covering distinct information requirements does not require finding every supporting document.

On TRQA, both Qwen and Gemma retrieve a larger share of qrel documents and ground more hops than Sol, yet answer fewer questions correctly (b1 vs.\ b2--b3).
Sol achieves higher P@all on every dataset; on TRQA, it reaches 70.9\%, compared with under 60\% for Qwen and Gemma.
\cmass{} makes these differences visible, allowing researchers to assess information diversity alongside document recall, precision, and answer accuracy.
These retrieval measures do not establish whether the agent used the evidence to answer.

\paragraph{Supplying evidence resolves incorrect answers.}
Supplying qrel documents directly in context, with tools and web access disabled, lets Sol correctly answer 32 of the 34 questions it missed in retrieval-enabled runs.
Human verification confirms Sol's remaining two errors despite the supporting evidence.
This sharp recovery points to evidence acquisition and presentation as promising routes to stronger agents, although it does not distinguish missed evidence from errors in reasoning over evidence retrieved in the original run.
Appendix~\ref{app:supplementary-results} gives the per-dataset results and evidence-supply procedure for this retrieval-augmented generation (RAG) setting.

These four findings together define concrete targets for agent training: covering missing hops, completing answer sets, and maintaining accuracy with less search effort.
Used as a held-out evaluation set, \cmass{} can test whether fine-tuned agents, query generators, or rerankers improve these outcomes across tasks over the same corpus.
For query generators and rerankers, this includes preserving hop coverage with fewer retrieved documents.
Closed-book runs check for accuracy gains \mbox{without retrieval}.

\section{Projection Challenges}
\label{sec:projection-challenges}
\paragraph{Exact answer sets and uniqueness.}
Set completeness and single-answer uniqueness require grounded exclusions of plausible alternatives.
For a set, every qualifying member must also be supported; for a single answer, the accepted reasoning chain must distinguish it from the alternatives.
Aggregates additionally require correctly identifying every contributing member and value: matching the source answer is insufficient because different or incomplete sets can yield the same number.
To satisfy these strict requirements, approximately 35\% of retained hops in the TRQA aggregation and DSQA set-answer subsets verify exclusions (Table~\ref{tab:hop-role-rates}).
Unresolved membership or exclusion causes projection failure.

\paragraph{Source questions and answers.}
Audits found typos and broken wording in \hotpot{}, filters that reveal the answer in \trqa{}, and inaccurate identifying clues in \bcp{}.
Some answers became stale over time; other errors and incomplete sets were present from the outset.
For \hotpot{} and Wikipedia-based \trqa{}, source evidence can omit qualifying entities or attributes, so broader web evidence may change an answer or count without any real-world change.
Projection therefore requires independent verification of source answers.
Table~\ref{tab:projection-challenge-examples} documents examples, repairs, and question exclusions due to failed projection.

\paragraph{Missing and conflicting evidence.}
Even this large corpus can lack evidence for a birth month, historical value, or attribution to a named source.
Recovering the supplied answer therefore does not guarantee support for every constraint or member of a gold answer set.
A known member cannot be omitted because its evidence is missing from the corpus.
When documents disagree, the audit checks dates, entity identities, definitions, and editions to determine whether the conflict changes the answer.
Permitted scope or constraint revisions can restore answerability but require renewed auditing; unresolved questions are excluded.

\paragraph{Agent reliability and reproducibility.}
Agents sometimes omitted constraints, accepted partial matches, or overlooked alternatives.
Context limits and session restarts also affected execution.
Full-document inspection, quotation checks, adversarial verification, and author review reduce these errors, but rerunning agents can still change evidence and acceptance decisions despite the fixed corpus.
Saved decisions and scripted validation support inspection and reconstruction of the pipeline outputs.
Appendix~\ref{app:revised-pipeline} details repeated auditing and human review; Appendix~\ref{app:challenges} shows how repairs and exclusions enforce these criteria.

\section{Related Work}
\label{sec:related}
Agentic search benchmarks differ in how they define the available evidence.
BrowseComp~\citep{wei2025browsecomp}, \dsqa{}~\citep{gupta2026deepsearchqa}, and LiveBrowseComp~\citep{fan2026livebrowsecomp} evaluate live-web search; the latter uses recent questions to reduce reliance on prior knowledge.
LoHoSearch~\citep{zhao2026lohosearch} constructs interdependent constraints in a Wikipedia-derived graph but evaluates agents through live-web search and browsing.
\bcp{}~\citep{chen2026browsecompplus} and OpenResearcher~\citep{li2026openresearcher} construct fixed corpora from task-specific supporting documents and distractors.
\cmass{} extends controlled evaluation to an independently assembled collection approximately $5{,}500\times$ the size of BCP's corpus and $37\times$ that of OpenResearcher, while broadening task and source coverage.

Earlier work projects web-derived answers onto a target corpus~\citep{Brill_etal_TREC2001} and transfers relevance judgments across corpora using query-specific classifiers~\citep{asadi2011crosscorpus}.
BrowseComp-Wikipedia~\citep{yang2026sira} selects BrowseComp questions supported by Wikipedia and evaluates retrieval of annotated answer-supporting pages.
NanoKnow~\citep{gu2026nanoknow} verifies answer passages for Natural Questions~\citep{kwiatkowski2019natural} and SQuAD~\citep{rajpurkar2016squad} in pre-training corpora to study parametric knowledge and supplied evidence.
\cmass{} verifies every constraint and reasoning hop across four datasets, including answer uniqueness, set completeness, and the members and values needed for aggregates.

Studies of retrieval interfaces motivate testing how findings transfer across corpora and tasks.
Pi-Serini~\citep{hsu2026piserini} studies configured BM25 within an agent on BCP, DCI~\citep{li2026dci} explores shell-based corpus access, and RISE~\citep{zhuang2026rise} uses BM25 to retrieve a bounded document set for shell-based exploration.
Concurrent with our work, Q2D-Web~\citep{schall2026q2dweb} evaluates retrievers using agent-reformulated queries, without measuring end-to-end answer correctness through iterative search.
\cmass{} supplies complete tasks and evidence annotations for comparing such approaches at larger scale; our present experiments establish BM25-based agent baselines.
Appendix~\ref{app:related-work} provides extended related work.

\FloatBarrier
\section{Conclusion and Future Work}

\cmass{} provides a shared setting for testing whether advances in agentic search generalize across tasks in a large, independently assembled corpus.
Its 423 verified QA pairs span four benchmarks grounded in 553M documents, with qrels supporting evaluation of both evidence retrieval and answer correctness.
Independent agent audits and human review verify grounding, uniqueness, and set completeness.

Our evaluations reveal substantial headroom for open-weight agents, including recovering complete answer sets, while high answer accuracy can coexist with incomplete evidence coverage.
Supplying evidence resolves 32 of GPT-6.1 Sol's 34 retrieval-enabled failures, motivating improvements in evidence acquisition and presentation.
Holding the evidence source fixed supports comparisons of retrieval methods and learned search strategies, with closed-book controls to measure gains beyond answering without search.

Fixed corpora can diverge from the live web over time; our corpus-agnostic pipeline supports projection onto newer collections with renewed verification of answers and qrels.
As benchmark questions enter training data, our dataset-agnostic pipeline can refresh the suite with new benchmarks such as LoHoSearch~\citep{zhao2026lohosearch}.

While our current focus is evaluation, projecting training datasets such as ORBIT~\citep{thakur2026orbit} onto \cm{} could support search-agent training without external search API costs.
ORBIT already links verification statements to source URLs, which should simplify projection relative to evaluation datasets without these annotations.

\section*{Limitations}
We retain only questions whose required evidence we can verify in \cm{}, and some retained questions undergo controlled revisions.
These choices can change the task mix and difficulty relative to the original datasets.

The datasets and corpus are English-only and text-only, leaving multilingual and multimodal search outside the current evaluation.
Extending to these settings requires suitable datasets, corpora, and verification methods.

Our audits and human review are best-effort checks and may miss evidence or make incorrect judgments, leaving qrels incomplete or erroneous.
Both retrieval metrics use pooled qrels.
Pooled judgments miss relevant documents~\citep{zobel1998reliable}, and pools that are small relative to the collection can be biased~\citep{buckley2007bias}, so our qrels can undercount valid evidence.
Because the pool includes documents retrieved by the three evaluated agents, it also favors them over new retrievers.
The literal hop-rule filter can miss relevant documents phrased differently from established evidence; unmatched pooled documents remain unjudged.
Duplicate expansion credits equivalent documents, but document recall remains sensitive to the number of supporting copies.

When answer uniqueness or set completeness relies on adversarial search, acceptance cannot rule out unseen qualifying candidates.
Guaranteeing that no qualifying candidate has been missed would require exhaustively checking every document in the corpus, which is infeasible at \cm{}'s scale of 553M documents.

Our projection pipeline uses BM25, while acceptance requires verified support for every factual hop, including confirmatory details and exclusions of plausible alternatives.
These strict quality requirements limit projection yield: a question can fail verification when retrieval misses evidence that is present in the corpus.
Stronger retrieval components, including dense or hybrid retrievers and rerankers, could surface more evidence and increase yield while preserving these quality checks.

\section*{Acknowledgments}
This research was supported in part by the Natural Sciences and Engineering Research Council of Canada (NSERC).

\bibliography{plusification}

@INPROCEEDINGS{Brill_etal_TREC2001,
   author = "Eric Brill and Jimmy Lin and Michele Banko and Susan Dumais and Andrew Ng",
   title = "Data-Intensive Question Answering",
   booktitle = "Proceedings of the Tenth Text REtrieval Conference (TREC 2001)",
   year = 2001,
   pages = "393--400",
   address = "Gaithersburg, Maryland",
}

@inproceedings{chen2026browsecompplus,
  title     = {{BrowseComp-Plus}: A Fair and Disentangled Evaluation Benchmark for Deep Search Agents},
  author    = {Chen, Zijian and Ma, Xueguang and Zhuang, Shengyao and Nie, Ping and Zou, Kai and Sharifymoghaddam, Sahel and Liu, Andrew and Green, Joshua and Patel, Kshama and Meng, Ruoxi and Su, Mingyi and Li, Yanxi and Hong, Haoran and Shi, Xinyu and Liu, Xuye and Oyarhoseini, Hosna and Thakur, Nandan and Zhang, Crystina and Gao, Luyu and Chen, Wenhu and Lin, Jimmy},
  booktitle = {Proceedings of the 64th Annual Meeting of the Association for Computational Linguistics (Volume 1: Long Papers)},
  pages     = {22349--22370},
  address   = {San Diego, California, United States},
  publisher = {Association for Computational Linguistics},
  doi       = {10.18653/v1/2026.acl-long.1023},
  url       = {https://aclanthology.org/2026.acl-long.1023/},
  year      = {2026}
}

@article{wei2025browsecomp,
  title   = {{BrowseComp}: A Simple Yet Challenging Benchmark for Browsing Agents},
  author  = {Wei, Jason and Sun, Zhiqing and Papay, Spencer and McKinney, Scott and Han, Jeffrey and Fulford, Isa and Chung, Hyung Won and Passos, Alex Tachard and Fedus, William and Glaese, Amelia},
  journal = {arXiv:2504.12516},
  year    = {2025}
}

@INPROCEEDINGS{DiaoShizhe_etal_NeurIPS2025,
   title = "{Nemotron-CLIMB}: Clustering-based Iterative Data Mixture Bootstrapping for Language Model Pre-training",
   author = "Shizhe Diao and Yu Yang and Yonggan Fu and Xin Dong and Dan Su and Markus Kliegl and Zijia Chen and Peter Belcak and Yoshi Suhara and Hongxu Yin and Mostofa Patwary and Yingyan (Celine) Lin and Jan Kautz and Pavlo Molchanov",
   booktitle = "Advances in Neural Information Processing Systems 38 (NeurIPS 2025) Datasets and Benchmarks Track",
   publisher = "Curran Associates, Inc.",
   year = 2025,
}

@inproceedings{Su:2412.02595:2024,
  title = "{Nemotron-CC}: Transforming {Common Crawl} into a Refined Long-Horizon Pretraining Dataset",
  author = "Dan Su and Kezhi Kong and Ying Lin and Joseph Jennings and Brandon Norick and Markus Kliegl and Mostofa Patwary and Mohammad Shoeybi and Bryan Catanzaro",
  booktitle = {Proceedings of the 63rd Annual Meeting of the Association for Computational Linguistics (Volume 1: Long Papers)},
  pages = {2459--2475},
  address = {Vienna, Austria},
  publisher = {Association for Computational Linguistics},
  doi = {10.18653/v1/2025.acl-long.123},
  url = {https://aclanthology.org/2025.acl-long.123/},
  year = {2025},
}

@inproceedings{gu2026nanoknow,
  author    = {Lingwei Gu and Nour Jedidi and Jimmy Lin},
  title     = {{NanoKnow}: How to Know What Your Language Model Knows},
  booktitle = {Proceedings of the 49th International {ACM} {SIGIR} Conference on Research and Development in Information Retrieval ({SIGIR} 2026)},
  year      = {2026},
  address   = {Melbourne, Australia},
  pages     = {3213--3219},
  publisher = {Association for Computing Machinery},
  doi       = {10.1145/3805712.3808604},
  url       = {https://doi.org/10.1145/3805712.3808604},
}

@article{kwiatkowski2019natural,
  author    = {Kwiatkowski, Tom and Palomaki, Jennimaria and Redfield, Olivia and Collins, Michael and Parikh, Ankur and Alberti, Chris and Epstein, Danielle and Polosukhin, Illia and Devlin, Jacob and Lee, Kenton and Toutanova, Kristina and Jones, Llion and Kelcey, Matthew and Chang, Ming-Wei and Dai, Andrew M. and Uszkoreit, Jakob and Le, Quoc and Petrov, Slav},
  title     = {{Natural Questions}: A Benchmark for Question Answering Research},
  journal   = {Transactions of the Association for Computational Linguistics},
  volume    = {7},
  pages     = {452--466},
  year      = {2019},
  doi       = {10.1162/tacl_a_00276},
  url       = {https://aclanthology.org/Q19-1026/}
}

@inproceedings{rajpurkar2016squad,
  author    = {Rajpurkar, Pranav and Zhang, Jian and Lopyrev, Konstantin and Liang, Percy},
  title     = {{SQuAD}: 100,000+ Questions for Machine Comprehension of Text},
  booktitle = {Proceedings of the 2016 Conference on Empirical Methods in Natural Language Processing},
  pages     = {2383--2392},
  year      = {2016},
  address   = {Austin, Texas},
  publisher = {Association for Computational Linguistics},
  doi       = {10.18653/v1/D16-1264},
  url       = {https://aclanthology.org/D16-1264/}
}

@misc{nanochat,
  author = {Andrej Karpathy},
  title = {nanochat: The best {ChatGPT} that \$100 can buy},
  year = {2025},
  publisher = {GitHub},
  howpublished = {\url{https://github.com/karpathy/nanochat}},
  url = {https://github.com/karpathy/nanochat}
}

@inproceedings{penedo2024fineweb,
  title     = {The {FineWeb} Datasets: Decanting the Web for the Finest Text Data at Scale},
  author    = {Penedo, Guilherme and Kydl{\'i}{\v{c}}ek, Hynek and Ben Allal, Loubna and Lozhkov, Anton and Mitchell, Margaret and Raffel, Colin and Von Werra, Leandro and Wolf, Thomas},
  booktitle = {Advances in Neural Information Processing Systems 37 (NeurIPS 2024) Datasets and Benchmarks Track},
  pages     = {30811--30849},
  publisher = {Curran Associates, Inc.},
  year      = {2024}
}

@article{thakur2026orbit,
  title   = {{ORBIT}: Scalable and Verifiable Data Generation for Search Agents on a Tight Budget},
  author  = {Thakur, Nandan and Chen, Zijian and Ma, Xueguang and Lin, Jimmy},
  journal = {arXiv:2604.01195},
  year    = {2026}
}

@inproceedings{rafiee2026trqa,
  title   = {{Total Recall QA}: A Verifiable Evaluation Suite for Deep Research Agents},
  author  = {Rafiee, Mahta and Soudani, Heydar and Abbasiantaeb, Zahra and Aliannejadi, Mohammad and Hasibi, Faegheh and Zamani, Hamed},
  booktitle = {Proceedings of the 49th International ACM SIGIR Conference on Research and Development in Information Retrieval (SIGIR 2026)},
  pages = {3365--3373},
  address = {Melbourne, Australia},
  publisher = {Association for Computing~Machinery},
  doi = {10.1145/3805712.3808629},
  url = {https://doi.org/10.1145/3805712.3808629},
  year    = {2026}
}

@article{gupta2026deepsearchqa,
  title   = {{DeepSearchQA}: Bridging the Comprehensiveness Gap for Deep Research Agents},
  author  = {Gupta, Nikita and Chatterjee, Riju and Haas, Lukas and Tao, Connie and Wang, Andrew and Liu, Chang and Oiwa, Hidekazu and Gribovskaya, Elena and Ackermann, Jan and Blitzer, John and Goldshtein, Sasha and Das, Dipanjan},
  journal = {arXiv:2601.20975},
  year    = {2026}
}

@article{wolfson2025monaco,
  title   = {{MoNaCo}: More Natural and Complex Questions for Reasoning Across Dozens of Documents},
  author  = {Wolfson, Tomer and Trivedi, Harsh and Geva, Mor and Goldberg, Yoav and Roth, Dan and Khot, Tushar and Sabharwal, Ashish and Tsarfaty, Reut},
  journal = {Transactions of the Association for Computational Linguistics},
  volume  = {14},
  pages   = {23--46},
  year    = {2026},
  doi     = {10.1162/tacl.a.64},
  url     = {https://aclanthology.org/2026.tacl-1.2/}
}

@inproceedings{krishna2024frames,
  title   = {Fact, Fetch, and Reason: A Unified Evaluation of Retrieval-Augmented Generation},
  author  = {Krishna, Satyapriya and Krishna, Kalpesh and Mohananey, Anhad and Schwarcz, Steven and Stambler, Adam and Upadhyay, Shyam and Faruqui, Manaal},
  booktitle = {Proceedings of the 2025 Conference of the Nations of the Americas Chapter of the Association for Computational Linguistics: Human Language Technologies (Volume 1: Long Papers)},
  pages = {4745--4759},
  address = {Albuquerque, New Mexico},
  publisher = {Association for Computational Linguistics},
  doi = {10.18653/v1/2025.naacl-long.243},
  url = {https://aclanthology.org/2025.naacl-long.243/},
  month = apr,
  year    = {2025}
}

@article{coelho2025deepresearchgym,
  title   = {{DeepResearchGym}: A Free, Transparent, and Reproducible Evaluation Sandbox for Deep Research},
  author  = {Coelho, Jo{\~a}o and Ning, Jingjie and He, Jingyuan and Mao, Kangrui and Paladugu, Abhijay and Setlur, Pranav and Jin, Jiahe and Callan, Jamie and Magalh{\~a}es, Jo{\~a}o and Martins, Bruno and Xiong, Chenyan},
  journal = {arXiv:2505.19253},
  year    = {2025}
}

@article{zhuang2026rise,
  title   = {Towards Retrieving Interaction Spaces for Agentic Search},
  author  = {Zhuang, Shengyao and Ni, Yuansheng and Fun, Hengxin and Lin, Jimmy and Ma, Xueguang},
  journal = {arXiv:2606.06880},
  year    = {2026}
}

@inproceedings{choi2026qcare,
  title   = {Towards Query-Agnostic {RAG} Evaluation via Query Coverage and Claim Verifiability},
  author  = {Choi, Jeonghwan and Yun, Taewon and Ban, Minjeong and Sun, Gyeonghun and Lee, Jae-Gil and Song, Hwanjun},
  booktitle = {Proceedings of the Conference on Language Modeling (COLM 2026)},
  year    = {2026}
}

@inproceedings{lins2021pyserini,
  title     = {Pyserini: A {Python} Toolkit for Reproducible Information Retrieval Research with Sparse and Dense Representations},
  author    = {Lin, Jimmy and Ma, Xueguang and Lin, Sheng-Chieh and Yang, Jheng-Hong and Pradeep, Ronak and Nogueira, Rodrigo},
  booktitle = {Proceedings of the 44th International ACM SIGIR Conference on Research and Development in Information Retrieval (SIGIR 2021)},
  pages     = {2356--2362},
  address   = {Virtual Event, Canada},
  publisher = {Association for Computing Machinery},
  year      = {2021}
}

@inproceedings{asadi2011crosscorpus,
  title     = {Cross-Corpus Relevance Projection},
  author    = {Asadi, Nima and Metzler, Donald and Lin, Jimmy},
  booktitle = {Proceedings of the 34th International ACM SIGIR Conference on Research and Development in Information Retrieval (SIGIR 2011)},
  pages     = {1163--1164},
  address   = {Beijing, China},
  publisher = {Association for Computing Machinery},
  year      = {2011}
}

@article{hsu2026piserini,
  title   = {Rethinking Agentic Search with {Pi-Serini}: Is Lexical Retrieval Sufficient?},
  author  = {Hsu, Tz-Huan and Yang, Jheng-Hong and Lin, Jimmy},
  journal = {arXiv:2605.10848},
  year    = {2026}
}

@article{sen2026grep,
  title   = {Is Grep All You Need? {How} Agent Harnesses Reshape Agentic Search},
  author  = {Sen, Sahil and Kasturi, Akhil and Lumer, Elias and Gulati, Anmol and Subbiah, Vamse Kumar},
  journal = {arXiv:2605.15184},
  year    = {2026}
}

@article{li2026dci,
  title   = {Beyond Semantic Similarity: Rethinking Retrieval for Agentic Search via Direct Corpus Interaction},
  author  = {Li, Zhuofeng and Zhang, Haoxiang and Wei, Cong and Lu, Pan and Nie, Ping and Lu, Yi and Bai, Yuyang and Feng, Shangbin and Zhu, Hangxiao and Zhong, Ming and Zhang, Yuyu and Xie, Jianwen and Choi, Yejin and Zou, James and Han, Jiawei and Chen, Wenhu and Lin, Jimmy and Jiang, Dongfu and Zhang, Yu},
  journal = {arXiv:2605.05242},
  year    = {2026}
}

@article{salemi2026grepseek,
  title   = {{GrepSeek}: Training Search Agents for Direct Corpus Interaction},
  author  = {Salemi, Alireza and Zeng, Chang and Nijasure, Atharva and Chung, Jui-Hui and Rahimi, Razieh and Diaz, Fernando and Zamani, Hamed},
  journal = {arXiv:2605.29307},
  year    = {2026}
}

@article{li2026openresearcher,
  title   = {{OpenResearcher}: A Fully Open Pipeline for Long-Horizon Deep Research Trajectory Synthesis},
  author  = {Li, Zhuofeng and Jiang, Dongfu and Ma, Xueguang and Zhang, Haoxiang and Nie, Ping and Zhang, Yuyu and Zou, Kai and Xie, Jianwen and Zhang, Yu and Chen, Wenhu},
  journal = {arXiv:2603.20278},
  year    = {2026}
}

@inproceedings{sun2025contextfolding,
  title   = {Scaling Long-Horizon Agent via Context Folding},
  author  = {Sun, Weiwei and Lu, Miao and Ling, Zhan and Liu, Kang and Yao, Xuesong and Yang, Yiming and Chen, Jiecao},
  booktitle = {Proceedings of the 43rd International Conference on Machine Learning},
  pages = {116894--116910},
  volume = {306},
  series = {Proceedings of Machine Learning Research},
  publisher = {PMLR},
  url = {https://proceedings.mlr.press/v306/sun26x.html},
  year    = {2026}
}

@article{chang2026karl,
  title   = {{KARL}: Knowledge Agents via Reinforcement Learning},
  author  = {Chang, Jonathan D. and Drozdov, Andrew and Toshniwal, Shubham and Oertell, Owen and Trott, Alexander and Portes, Jacob and Gupta, Abhay and Koppol, Pallavi and Baheti, Ashutosh and Kulinski, Sean and Zhou, Ivan and Dea, Irene and Opsahl-Ong, Krista and Favreau-Lessard, Simon and Owen, Sean and Gonzalez Ortiz, Jose Javier and Singhvi, Arnav and Andrade, Xabi and Wang, Cindy and Sreenivasan, Kartik and Havens, Sam and Liu, Jialu and DeNiro, Peyton and Sun, Wen and Bendersky, Michael and Frankle, Jonathan},
  journal = {arXiv:2603.05218},
  year    = {2026}
}

@article{yang2026sira,
  title   = {Superintelligent Retrieval Agent: The Next Frontier of Agentic Retrieval},
  author  = {Yang, Zeyu and Han, Xu and Ma, Qi and Chen, Jason and Shrivastava, Anshumali},
  journal = {arXiv:2605.06647},
  year    = {2026}
}

@article{fan2026livebrowsecomp,
  title = {{LiveBrowseComp}: Are Search Agents Searching, or Just Verifying What They Already Know?},
  author = {Fan, HuiMing and Wang, Xiao and Chu, Zheng and Wang, Qianyu and Wang, Zhuoyao and Liu, Ming and Qin, Bing and {XingYu}},
  journal = {arXiv:2605.28721},
  year = {2026}
}

@article{zhao2026lohosearch,
  title   = {{LoHoSearch}: Benchmarking Long-Horizon Search Agents Beyond the Human Difficulty Ceiling},
  author  = {Zhao, Jiarui and Zhang, Rongzhi and Liu, Lingchuan and Yang, Hao and Cai, Xunliang and Su, Xi},
  journal = {arXiv:2606.12837},
  year    = {2026}
}

@inproceedings{solodko2026lakequest,
  title     = {{LakeQuest}: A Three-Domain Benchmark for Grounded Question Answering across Data Lakes},
  author    = {Solodko, Michael and Gong, Steven and Yu, Guangwei and Gorti, Satya Krishna and Cresswell, Jesse C. and Zhong, Victor},
  booktitle = {Proceedings of the Conference on Language Modeling (COLM 2026)},
  year      = {2026}
}

@article{schall2026q2dweb,
  title = {{Q2D-Web}: A Large-Scale Benchmark for Retrieval in Agentic {RAG} Systems},
  author = {Schall, Maximilian and Eslami, Sedigheh and Krimmel, Markus and Chaffin, Antoine and Milliken, Louis and Wang, Bo and Bykov, Denis},
  journal = {arXiv:2609.08887},
  year = {2026}
}

@inproceedings{yang2018hotpotqa,
  title = {{HotpotQA}: A Dataset for Diverse, Explainable Multi-hop Question Answering},
  author = {Yang, Zhilin and Qi, Peng and Zhang, Saizheng and Bengio, Yoshua and Cohen, William and Salakhutdinov, Ruslan and Manning, Christopher D.},
  booktitle = {Proceedings of the 2018 Conference on Empirical Methods in Natural Language Processing},
  year = {2018},
  address = {Brussels, Belgium},
  publisher = {Association for Computational Linguistics},
  pages = {2369--2380},
  doi = {10.18653/v1/D18-1259},
  url = {https://aclanthology.org/D18-1259/}
}

@inproceedings{cormack2009rrf,
  title     = {Reciprocal Rank Fusion Outperforms {Condorcet} and Individual Rank Learning Methods},
  author    = {Cormack, Gordon V. and Clarke, Charles L. A. and B{\"u}ttcher, Stefan},
  booktitle = {Proceedings of the 32nd International ACM SIGIR Conference on Research and Development in Information Retrieval},
  pages     = {758--759},
  address   = {Boston, MA, USA},
  publisher = {Association for Computing Machinery},
  doi       = {10.1145/1571941.1572114},
  year      = {2009}
}

@techreport{sparckjones1975ideal,
  title       = {Report on the Need for and Provision of an ``Ideal'' Information Retrieval Test Collection},
  author      = {Sp{\"a}rck Jones, Karen and van Rijsbergen, C. J.},
  institution = {Computer Laboratory, University of Cambridge},
  type        = {British Library Research and Development Report},
  number      = {5266},
  year        = {1975}
}

@book{voorhees2005trec,
  title     = {{TREC}: Experiment and Evaluation in Information Retrieval},
  editor    = {Voorhees, Ellen M. and Harman, Donna K.},
  publisher = {MIT Press},
  address   = {Cambridge, MA},
  series    = {Digital Libraries and Electronic Publishing},
  isbn      = {0262220733},
  year      = {2005}
}

@inproceedings{zobel1998reliable,
  title     = {How Reliable Are the Results of Large-Scale Information Retrieval Experiments?},
  author    = {Zobel, Justin},
  booktitle = {Proceedings of the 21st Annual International ACM SIGIR Conference on Research and Development in Information Retrieval},
  pages     = {307--314},
  address   = {Melbourne, Australia},
  publisher = {Association for Computing Machinery},
  doi       = {10.1145/290941.291014},
  year      = {1998}
}

@article{buckley2007bias,
  title   = {Bias and the Limits of Pooling for Large Collections},
  author  = {Buckley, Chris and Dimmick, Darrin and Soboroff, Ian and Voorhees, Ellen},
  journal = {Information Retrieval},
  volume  = {10},
  number  = {6},
  pages   = {\mbox{491--508}},
  doi     = {10.1007/s10791-007-9032-x},
  year    = {2007}
}

\clearpage
\appendix
\renewcommand{\floatpagefraction}{0.95}
\renewcommand{\dblfloatpagefraction}{0.95}
\renewcommand{\dbltopfraction}{0.95}
\renewcommand{\textfraction}{0.05}
\setcounter{dbltopnumber}{3}
\makeatletter
\setlength{\@fptop}{0pt}
\setlength{\@dblfptop}{0pt}
\setlength{\@fpsep}{12pt plus 2pt minus 2pt}
\setlength{\@dblfpsep}{12pt plus 2pt minus 2pt}
\makeatother

\section{Version History}
\label{app:version-history}
\cmass{} is an expanding suite, and this history records the scope of each manuscript revision and its accompanying data and code releases.
The present revision describes \cmass{} v1.0.

\paragraph{20 August 2026: Initial BCP projection.}
The initial manuscript introduced a pipeline for projecting BCP onto \cm{} by decomposing questions into reasoning hops, grounding each hop, and validating the evidence through independent agent checks and human review.
It evaluated retrieval and answer accuracy on 57 grounded BCP questions while preserving the source questions and answers.
The accompanying release provided those questions and answers, question-level qrels, projection code, and corpus-analysis artifacts, including exact and near-duplicate groups.

\paragraph{25 October 2026: CMASS v1.0.}
This revision expands the suite to BCP, TRQA, DSQA, and HotpotQA and adds controlled QA revision and adversarial checks of answer uniqueness and set completeness.
It reports evaluations of three models with retrieval and closed-book controls, measurements of hop groundedness, and supplied-evidence diagnostics.
The accompanying release is prepared as \texttt{v2026.10.25}, with 423 QAs and 40{,}736 qrels, inclusion/exclusion annotations, structured DSQA answer sets, and updated pipeline and evaluation code.
Qrels construction combines audited evidence, a recall pass, documents pooled from the evaluated agents, and duplicate expansion.
Table~\ref{revised:tab:yield} gives the per-dataset counts.
Reported results should identify the dataset version used.

\section{Extended Related Work}
\label{app:related-work}
\label{app:wiki-limitations}
\paragraph{Task requirements.}
Search benchmarks differ in what an agent must establish before its answer is complete.
BrowseComp, HotpotQA, FRAMES, and MoNaCo test difficult fact finding and reasoning across documents~\citep{wei2025browsecomp,yang2018hotpotqa,krishna2024frames,wolfson2025monaco}.
DSQA and TRQA add exhaustive answer sets and aggregation~\citep{gupta2026deepsearchqa,rafiee2026trqa}: finding one qualifying entity is insufficient when the question asks for every member or a calculation over them.
\cmass{} brings these tasks into a shared corpus to test whether search methods generalize across answer types.

\paragraph{Corpus construction and answerability.}
Using a fixed corpus ensures that agents can search the same documents across evaluation runs; it does not determine how those documents were selected.
BCP's evaluation corpus and OpenResearcher's trajectory-synthesis corpus both include supporting documents selected for their task questions, alongside distractors~\citep{chen2026browsecompplus,li2026openresearcher}.
DeepResearchGym instead supports reproducible retrieval and report evaluation over existing web collections, including FineWeb~\citep{coelho2025deepresearchgym,penedo2024fineweb}.
\cmass{} uses \cm{}, a pre-training corpus assembled independently of the questions, and establishes answerability by auditing each question's constraints and answer against the available evidence.
No external documents are added to make a question answerable.
Questions without sufficient verified support are revised within specified limits or excluded.

The target collection also determines which tasks can be transferred.
TRQA's synthetic commerce tasks and LakeQuest's domain-specific records require their underlying data and schemas~\citep{rafiee2026trqa,solodko2026lakequest}; their evidence cannot be assumed present in a general web crawl.
\cmass{} focuses on questions grounded in general web text, across a broader range of sources than Wikipedia alone.

\paragraph{Projection and answer completeness.}
Moving a task to another corpus requires more than retaining its question and answer.
Earlier work locates target-corpus evidence for answers found on the web~\citep{Brill_etal_TREC2001} or transfers relevance judgments between corpora~\citep{asadi2011crosscorpus}.
BrowseComp-Wikipedia and NanoKnow likewise select existing questions by checking answer support in Wikipedia and pre-training corpora, respectively~\citep{yang2026sira,gu2026nanoknow}.
\cmass{} verifies the full task: every constraint and reasoning hop, exclusions needed to establish uniqueness or set completeness, and the members and values used in aggregates.

\paragraph{Evidence retrieval and answer evaluation.}
After grounding the questions, evaluating a retriever and evaluating a search agent remain different tasks.
Concurrent with our work, Q2D-Web evaluates first-stage retrieval for agent-reformulated queries, using a corpus pooled from production retrieval results~\citep{schall2026q2dweb}.
\cmass{} instead retains the complete task: the agent formulates queries, searches repeatedly, and produces an answer.
Document recall and hop coverage can therefore be compared with final-answer correctness.
This distinction between finding evidence and using it is also reflected in Q-CARE's separate measures of sub-query coverage and claim verifiability~\citep{choi2026qcare}.
\looseness=-1

\paragraph{Generalization of search methods.}
\cmass{} combines varied tasks and evidence annotations to test whether search improvements transfer across answer types and corpus scale.
Experiments with configured BM25, direct corpus access through shell tools, and tool-result presentation show that effectiveness depends on the retrieval interface as well as the model~\citep{hsu2026piserini,li2026dci,sen2026grep}.
Larger collections introduce additional costs: DCI reports declining accuracy and rising costs as distractors are added~\citep{li2026dci}.
Retrieving a bounded set of documents for shell exploration and training agents to search corpus files are complementary approaches to corpus access~\citep{zhuang2026rise,salemi2026grepseek}.
\cmass{} provides a larger independent collection for evaluating such approaches.

For learned search strategies, interpreting benchmark results also requires knowing which data were used for training.
BCP-based reinforcement learning uses either questions as-is or synthetic questions generated from BCP questions and corpus documents~\citep{sun2025contextfolding,chang2026karl}.
\cmass{} extends evaluation across four source datasets in a shared retrieval environment, allowing learned strategies to be assessed beyond a single source benchmark.

Correct answers can also come from prior knowledge rather than successful search.
LiveBrowseComp addresses this issue with questions about recent, less prominent facts~\citep{fan2026livebrowsecomp}, but a fixed corpus collected before those facts cannot provide their evidence.
\cmass{} keeps the corpus fixed and reports closed-book accuracy alongside retrieval-enabled accuracy and evidence coverage, making the contribution of search an explicit part of the evaluation.

\section{Human Review Interface}
\label{app:review-ui}
\begin{figure*}[!tp]
\centering
\includegraphics[width=\textwidth,height=0.82\textheight,keepaspectratio]{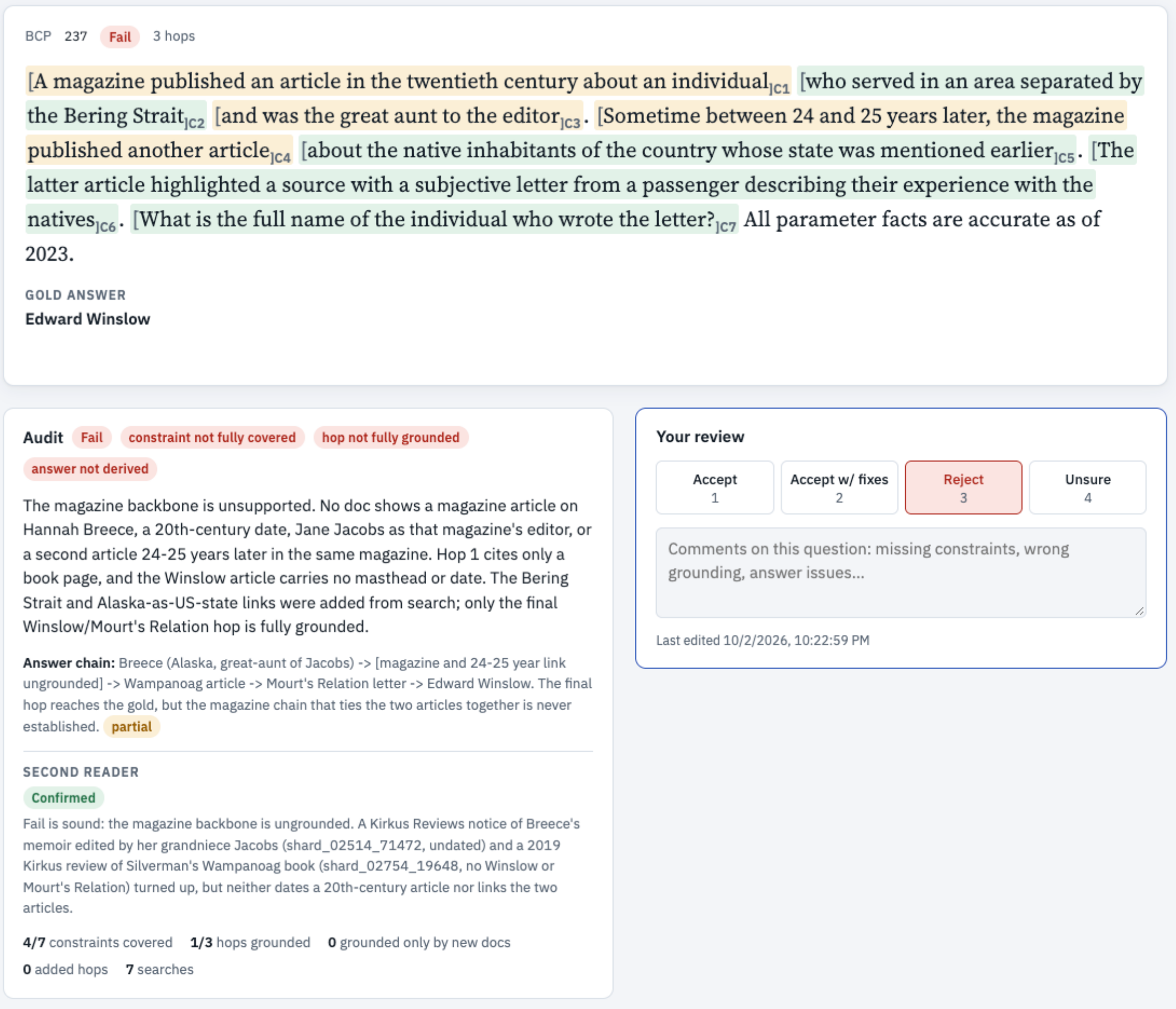}
\caption{Question-level audit and author review for BCP QID~237.
The interface displays the question, answer, audit findings, coverage summary, and human review decision.}
\label{fig:review-ui-overview}
\end{figure*}

\begin{figure*}[!tp]
\centering
\includegraphics[width=\textwidth,height=0.82\textheight,keepaspectratio]{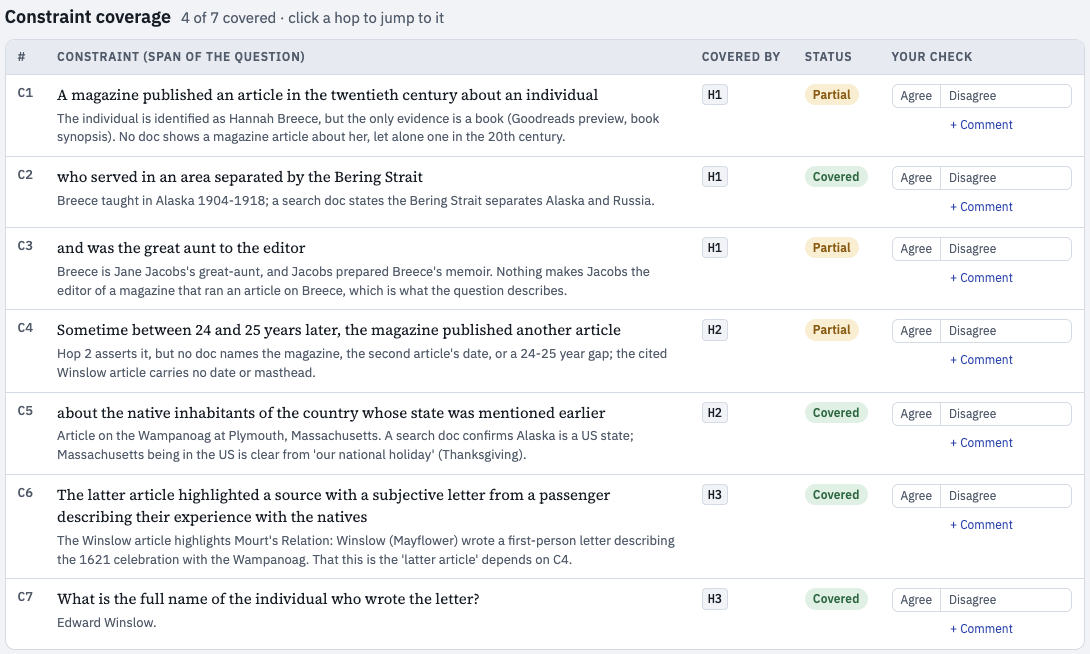}
\caption{Constraint coverage for BCP QID~237.
Each question span is linked to its supporting hops and a coverage status; reviewers can agree, disagree, or add a comment.}
\label{fig:review-ui-constraints}
\end{figure*}

\begin{figure*}[!tp]
\centering
\includegraphics[width=\textwidth,height=0.82\textheight,keepaspectratio]{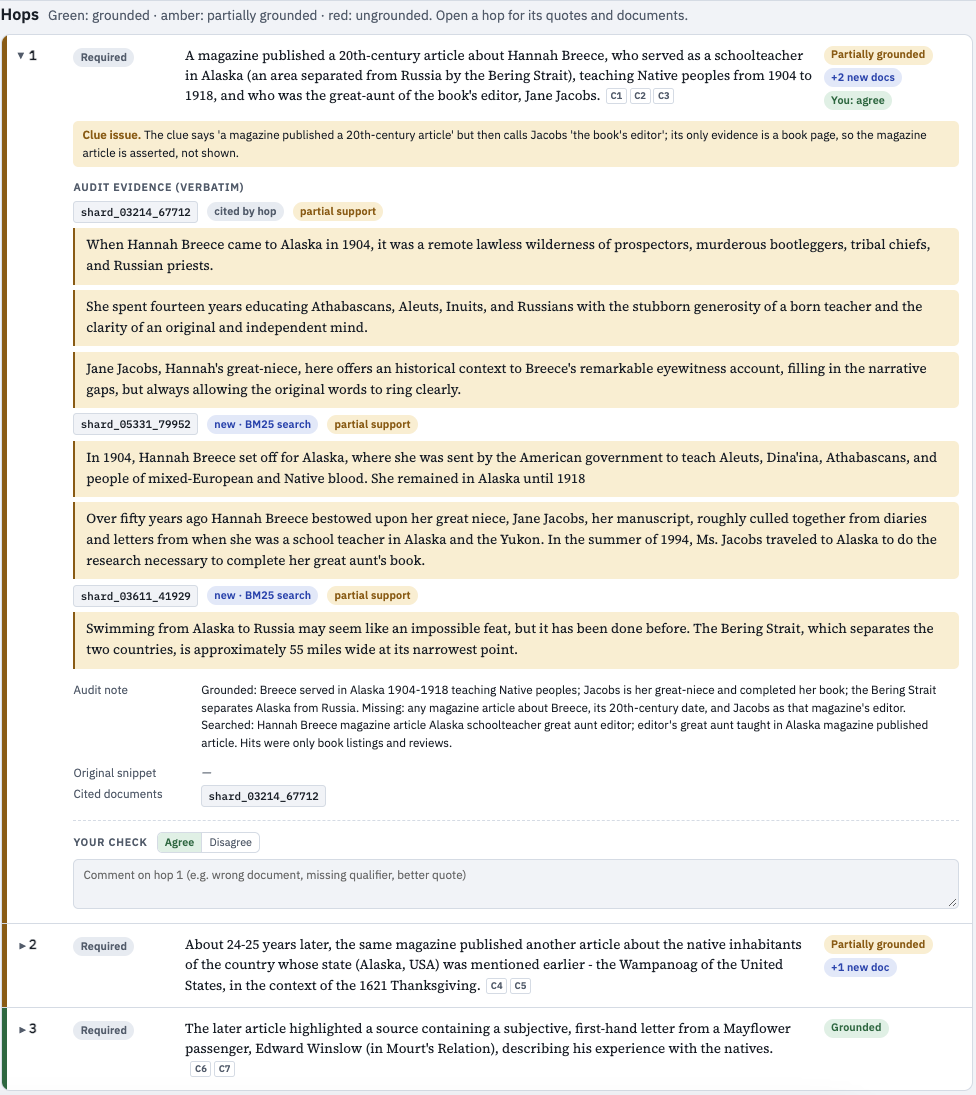}
\caption{Hop-level evidence for BCP QID~237.
An expanded hop shows document identifiers, quotations, grounding assessments, and audit notes, with controls for the reviewer's judgment and comments.}
\label{fig:review-ui-hops}
\end{figure*}

One author inspects agent audit findings and records review decisions through the interface.
Figures~\ref{fig:review-ui-overview}--\ref{fig:review-ui-hops} show three views of the same BCP question (QID~237): the question-level audit and review decision, constraint coverage, and hop evidence.
The reviewer can inspect the links between question spans, hops, and supporting quotations, and comment on individual constraints or hops.
Appendix~\ref{app:revised-pipeline} specifies how these decisions enter the audit and revision loop.

\section{Audit, Revision, and Human Review}
\label{app:revised-pipeline}

\begin{figure*}[!tp]
\centering
\includegraphics[width=\textwidth]{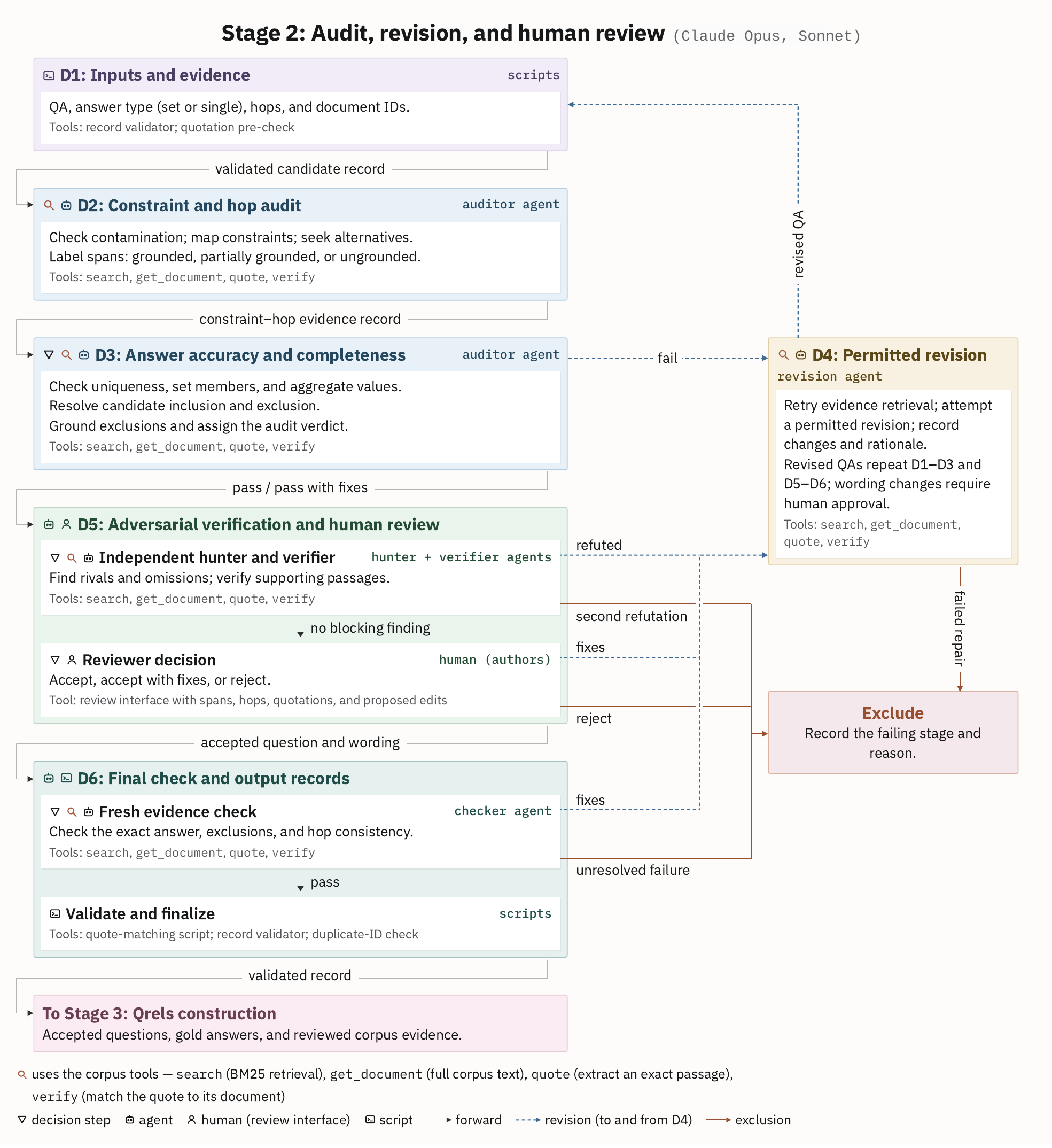}
\caption{Stage~2's audit, revision, and human verification loop, with D1--D6 corresponding to the subsections of Appendix~\ref{app:revised-pipeline}.
The figure identifies the agents, human reviewers (authors), scripts, and tools used at each step.
Dashed blue arrows route revisions through D4; revised QAs repeat the audit and review sequence.
Red arrows lead to exclusion.
HotpotQA skips the post-review checker agent and proceeds directly to script validation.}
\label{revised:fig:stage3}
\end{figure*}

Stage~2 audits question--answer (QA) candidates from all four \cmass{} datasets before their accepted evidence is used for Stage~3 qrels construction.
Candidates include both Stage~1 projections and near misses with unresolved evidence or answerability gaps.
An auditor checks the question and answer, a revision agent attempts permitted repairs, independent hunter and verifier agents challenge the result, and one author makes the human acceptance decision.
Figure~\ref{revised:fig:stage3} shows the six workflow steps, D1--D6, described below.
The following requirements apply throughout the workflow, including after revision.

\paragraph{Grounding and consistency.}
A \emph{hop} is an atomic reasoning step used to identify an answer, compute its value, or exclude an alternative.
Every factual constraint and factual hop must be supported by \cm{} evidence read in its full-document context.
The evidence must establish the intended entity, relation, value, and qualifiers; matching names alone do not establish identity.
Date restrictions require evidence for the specified period, and a clause such as ``according to X'' requires attribution to that source.
Hops must agree on entity identity, edition, date, unit, and scope; conflicts that affect the answer remain unresolved until the evidence settles them.

Evidence supplied by the source benchmark, model knowledge, and live-web results may guide searches but cannot replace corpus evidence.
Computations must use grounded inputs.
The human reviewer may explain logical inferences from grounded premises, but these explanations are distinguished from quotations and cannot supply missing facts.

\paragraph{Answer validity, uniqueness, and completeness.}
A single answer must be uniquely supported; a set must contain every qualifying member and no non-member; an aggregate additionally requires the correct values and computation.
Any requested ordering must also be verified.
Every plausible answer candidate found during the audit must be supported as qualifying or excluded by a grounded fact showing which constraint it fails.
Retrieval failure establishes neither inclusion nor exclusion.

Set completeness can rest on a corpus document giving the full list or its size, or on a fixed domain whose members are all checked.
Otherwise, acceptance requires a recorded search that finds no additional qualifying member and leaves no plausible candidate unresolved; D2 specifies the search procedure.
This search-based judgment does not prove that retrieval has exhausted the corpus.

\paragraph{Question quality and audit failures.}
False premises, malformed wording, ambiguity, and incorrect or stale answers are flagged for repair rather than interpreted in favor of the supplied answer.
Revisions must preserve a nontrivial question.
Violating any shared requirement prevents a passing audit verdict.
A question is excluded if a single corpus document reproduces both the question and answer or supplies every fact needed to answer it.
Other failures enter revision under D4's limits.

\subsection{Inputs and Evidence}
\label{revised:sec:where}
At D1, a script validates the candidate record: source question and answer, recorded answer type (set or single), proposed hops, supporting document identifiers, and snippets.
Agents use \texttt{search} for BM25 retrieval, \texttt{get\_document} for full corpus text, and \texttt{quote} to extract passages.
The \texttt{verify} script checks that quotations occur in their cited documents, ignoring differences in spaces, tabs, and line breaks.
Matching a quotation confirms its presence; D2 judges whether it supports the claimed fact.
Queries, document identifiers, and quotations are saved with the record.

\subsection{Constraint and Hop Audit}
\label{revised:sec:substeps}
At D2, the auditor derives the question's constraints from its wording and treats the proposed decomposition as provisional.
Constraints are recorded as non-overlapping spans covering entity descriptions, relations, dates, quantities, comparisons, exclusions, source restrictions, and the requested attribute.
The auditor links each span to the hops that establish it, retrieving evidence and adding omitted hops as needed.
Several hops may establish one constraint, and one hop may support several constraints.

Constraints and factual hops receive \emph{grounded}, \emph{partially grounded}, or \emph{ungrounded} labels according to whether the evidence establishes all, some, or none of their requirements.
Computational hops receive \emph{derivation only} labels and record their inputs and operation.
Each factual hop records supporting quotations and document identifiers or identifies the missing evidence.

To find rivals and missing members, the auditor searches list documents, known candidates by name, and category descriptions, inspecting each query's top 50 hits.
The evidence about rivals and set members accompanies the constraint--hop mapping sent to D3.
Auditor prompts include examples of reasoning steps rejected during human review of BCP questions outside the audit set.
These illustrate errors such as unsupported dates and missing facts needed to connect two reasoning steps.

\subsection{Answer Accuracy and Completeness}
\label{revised:sec:standards}
At D3, the auditor uses D2's record to reconstruct the answer, determine which potential answers qualify, and recompute any aggregate under the shared audit requirements.
It checks the result against the proposed answer and records any mismatch or unresolved issue.

\label{revised:sec:records}
The verdict is \emph{pass} if no changes are needed, \emph{pass with fixes} if adjusting hops or grounding suffices without changing the question or answer, and \emph{fail} otherwise.
All such adjustments are completed before a passing record advances to D5; failures go to D4.
These are agent audit verdicts, distinct from the author's acceptance decisions at D5.

\subsection{Permitted Revision}
\label{revised:sec:repairs}
D4 handles D3 failures and repair requests from adversarial verification, human review, or the final check.
The revision agent first retries retrieval and rereads the relevant documents, even if an earlier agent already searched for the missing evidence.
If evidence alone resolves the issue, the question and answer remain unchanged.
Otherwise, the agent attempts the least extensive permitted edit: change the wording while preserving the answer, recompute the answer without changing the wording, or revise both.
Table~\ref{revised:tab:adapters} defines the dataset-specific limits.
BCP's only source-answer change corrects a typo.
The record preserves the source QA and records added evidence, hop adjustments, wording or answer changes, and their rationale.

An unsuccessful repair excludes the candidate.
Every repaired candidate returns to D1 and repeats D2--D3, adversarial verification and human review at D5, and the applicable D6 checks.
Human approval covers the final wording and evidence; approval of an earlier version does not carry over.
If human review changes a general audit rule, earlier records affected by that rule are reopened.
D5 and D6 specify the additional retry limits for failures found at those steps.

\begin{table*}[t]
\tablestyle
\centering
\begin{tabularx}{\textwidth}{@{}lL@{}}
\toprule
Dataset & Permitted revisions \\
\midrule
\hotpot & Correct malformed wording or add one grounded qualifier; preserve the answer. \\
\bcp & Evidence-only repairs for questions retained by Stage~1; near misses may receive a clue correction or removal, a range adjustment, or one grounded qualifier; preserve the answer except for typo corrections. \\
\trqa & Add a grounded qualifier or change the requested attribute when the wording makes the answer trivial; recompute the answer only under the completeness requirements. \\
\dsqa & Remove a source clause, add a grounded qualifier, or adjust a year or threshold; recompute the answer only under the completeness requirements. \\
\bottomrule
\end{tabularx}
\caption{Dataset-specific limits on question and answer revisions.}
\label{revised:tab:adapters}
\end{table*}

\subsection{Adversarial Verification and Human~Review}
\label{revised:sec:flow}
At D5, an independent hunter challenges each passing candidate by searching for rival answers, missing or wrongly included members, conflicting values, and alternative readings of the question.
Hunters request up to 50 BM25 results per query, reformulate searches, and can use live-web search to identify leads.
For aggregates and changed answers, two hunters pursue complementary routes: list documents and named candidates.
A separate verifier reads the cited passages in context and decides whether each finding invalidates the proposed QA under the shared requirements.
A verified refutation permits one additional repair attempt at D4; a second refutation excludes the candidate.

When no blocking finding remains, the author inspects the question and answer, constraint--hop mapping, quotations, candidate decisions, and adversarial findings in the review interface (Figures~\ref{fig:review-ui-overview}--\ref{fig:review-ui-hops}).
Source and revised wording are displayed side by side, and the reviewer can comment on individual constraints or hops.
They record \emph{accept}, which advances the candidate to D6; \emph{accept with fixes}, which returns it to D4; or \emph{reject}, which excludes it.

\subsection{Final Check and Output Records}
\label{revised:app:checklist}
After human acceptance, \bcp{}, \trqa{}, and \dsqa{} receive a fresh agent check because their permitted revisions can remove constraints or change answers.
\hotpot{} proceeds directly to record validation.
The checker rereads the cited documents and searches for omissions, applying the shared requirements to the accepted QA and checking that any departure from the source answer has an approved rationale.
A grounding gap permits one evidence-only repair attempt at D4; a problem requiring new wording follows D4's dataset-specific revision limits.
Unresolved failures are excluded with the checker's reasons.

All retained records then undergo script validation: quotation matching is repeated, a failed match stops finalization, and duplicate candidates from multiple input lists are included only once.
The final record contains the source and revised QA, edit rationale, constraint--hop mapping, quotations, reviewer inference notes, candidate decisions, search history, audit verdict, human decision, and final-check outcome or exemption.
Set-answer records preserve members, aliases, values, and requested ordering; DSQA evaluation ignores order.
Excluded records retain the failure stage and reason.
Accepted records supply the supporting documents for Stage~3 qrels construction.

\section{Qrels Construction Details}
\label{app:construction-details}
\label{app:stage3-details}

\paragraph{Recall pass.}
For each factual hop, the preparation agent writes four to eight queries and anchor strings.
Each query retrieves 100 BM25 results; rankings are fused with reciprocal rank fusion ($k=60$), filtered by the anchors, and cut to the top 40 documents per hop for \bcp{} and 20 otherwise.
Judges assess each candidate against all live hops of its question, not only the hop whose query retrieved it.
\looseness=-1

\paragraph{Second judge and adjudication.}
Across the initial candidate pool and system pool, Claude Sonnet~5.5 independently re-judges 23{,}096 of 48{,}991 judged question--document pairs (47.1\%) without seeing the first verdict.
The first and second judges disagree on whether to retain the document in 2{,}043 of these 23{,}096 reviews (8.8\%).
A separate Claude Sonnet~5.5 adjudicator resolves these disagreements.
\looseness=-1

\paragraph{System pool.}
The retrieval logs of GPT-6.1 Sol, Gemma~4 31B IT, and Qwen~3.5 9B on all 423 questions list every document their searches returned or that they opened, without ranks, so a depth cut as in TREC is not possible.
Of the 49{,}885 question--document pairs in their union, 41{,}312 were neither qrels nor judged earlier.

For each hop, an agent writes match rules: clauses of alternative phrases (the hop's subject names and aliases, the fact's key terms), matched case-insensitively on word boundaries, where a document must contain one phrase from every clause.
The rules of a hop must match all of the hop's previously established qrels.
In total, 6{,}036 rules cover 2{,}693 hops.

A pooled document is judged only if it matches a rule of its question: 4{,}403 pairs, judged with the same standards and tiers as above, without the judges seeing which system retrieved them or which rule matched.
Documents that fail the rule filter remain unjudged.
\Cref{tab:stage3-sources} reports the counts, and \Cref{tab:stage3-systems} credits every qrel added by the system pool to the systems that retrieved it.
\looseness=-1

\paragraph{Example of anchor and rule filters.}
Consider the synthetic hop ``The Lumen Observatory was founded in 1987.''
For the recall pass, anchor strings could be \texttt{Lumen Observatory} and \texttt{Lumen Obs.}; a candidate passes the anchor filter if either string occurs in its text, ignoring case and normalizing whitespace.

A corresponding system-pool rule could require three clauses: (\texttt{Lumen Observatory} OR \texttt{Lumen Obs.}) AND (\texttt{founded} OR \texttt{established}) AND \texttt{1987}.
The rule matches ``The Lumen Obs. was established in 1987,'' but not a document mentioning only the observatory's name.
A match selects the document for judgment; the judge still verifies that it supports the hop, since the terms can occur without expressing the required relation.
\looseness=-1

\begin{table*}[t]
\tablestyle
\centering
\begin{tabular}{@{}lrrrrr@{}}
\toprule
 & \bcp{} & \trqa{} & \dsqa{} & \hotpot{} & Total \\
\midrule
Questions & 80 & 68 & 71 & 204 & 423 \\
\midrule
\multicolumn{6}{@{}l}{\emph{Stage 3 judgments}} \\
Candidates & 13{,}484 & 9{,}629 & 11{,}645 & 13{,}400 & 48{,}158 \\
\quad Stage~2 evidence (kept, not judged) & 861 & 784 & 1{,}183 & 742 & 3{,}570 \\
\quad Judged & 12{,}623 & 8{,}845 & 10{,}462 & 12{,}658 & 44{,}588 \\
\quad Re-judged by second judge & 7{,}018 & 3{,}705 & 4{,}887 & 4{,}454 & 20{,}064 \\
\quad Adjudicated & 592 & 261 & 451 & 434 & 1{,}738 \\
\textbf{Qrels before duplicate expansion} & \textbf{10{,}385} & \textbf{7{,}151} & \textbf{7{,}969} & \textbf{9{,}410} & \textbf{34{,}915} \\
\textbf{Duplicates added} & \textbf{583} & \textbf{361} & \textbf{729} & \textbf{692} & \textbf{2{,}365} \\
\midrule
\multicolumn{6}{@{}l}{\emph{System pool}} \\
Pooled pairs not yet judged & 20{,}005 & 3{,}135 & 7{,}849 & 10{,}323 & 41{,}312 \\
\quad Matching a hop rule (judged) & 758 & 950 & 1{,}579 & 1{,}116 & 4{,}403 \\
\quad \textbf{Kept} & \textbf{441} & \textbf{728} & \textbf{1{,}046} & \textbf{887} & \textbf{3{,}102} \\
\textbf{Duplicates added} & \textbf{41} & \textbf{67} & \textbf{145} & \textbf{101} & \textbf{354} \\
\midrule
Final qrels & 11{,}450 & 8{,}307 & 9{,}889 & 11{,}090 & 40{,}736 \\
\bottomrule
\end{tabular}
\caption{Qrel construction statistics by dataset.
Bold rows show the contributions that sum to the final qrel counts.}
\label{tab:stage3-sources}
\end{table*}

\begin{table}[!ht]
\tablestyle
\setlength{\tabcolsep}{3.5pt}
\centering
\begin{tabular}{@{}lrrrrr@{}}
\toprule
Retrieved by & \bcp{} & \trqa{} & \dsqa{} & \hotpot{} & Total \\
\midrule
Sol only & 171 & 101 & 194 & 132 & 598 \\
Gemma only & 100 & 309 & 318 & 278 & 1{,}005 \\
Qwen only & 162 & 258 & 542 & 391 & 1{,}353 \\
Sol + Gemma & 9 & 31 & 34 & 37 & 111 \\
Sol + Qwen & 18 & 16 & 26 & 17 & 77 \\
Gemma + Qwen & 10 & 71 & 62 & 109 & 252 \\
All three & 12 & 9 & 15 & 24 & 60 \\
\midrule
Total & 482 & 795 & 1{,}191 & 988 & 3{,}456 \\
\bottomrule
\end{tabular}
\caption{Qrels added by the system pool, by the agents that retrieved them.
A duplicate counts for the agents that retrieved the pooled document it copies.}
\label{tab:stage3-systems}
\end{table}

\section{Sample Issues Identified by the Audit}
\label{app:challenges}

Table~\ref{tab:projection-challenge-examples} presents sample issues identified by the audit and review process in Appendix~\ref{app:revised-pipeline}, including errors in source questions and answers and missing or conflicting evidence in \cm{}.
Failure to find evidence leaves an answer unverified; it does not show that the answer is wrong.
When documents report different values, the audit checks their context, including the entity and time period, to determine which value the question requires.

Each example records the correction made or why the question was excluded, illustrating how the process enforces the requirements for grounded, unambiguous, and complete answers.
Section~\ref{sec:projection-challenges} explains how these issues affect projection and reproducibility.

\begin{table*}[!tp]
\tablestyle
\centering
\begin{tabularx}{\textwidth}{@{}>{\raggedright\arraybackslash}p{0.165\textwidth}>{\raggedright\arraybackslash}p{0.35\textwidth}>{\raggedright\arraybackslash}X@{}}
\toprule
Issue, dataset, and QID & Original task or answer & Audit finding and outcome \\
\midrule
Broken wording\newline HotpotQA\newline {\nolinkurl{388}} &
The man associated with the award won by \emph{Summer of the Monkeys} was a spokesman for ``white sector of America.'' &
The question omits the relation ``is named after'' and misspells ``what'' as ``white.''
Both spans were repaired; the answer remains ``middle America.'' \\
\addlinespace[3pt]
Trivial question\newline TRQA\newline {\nolinkurl{wiki2__quest_281_p577}} &
Find the earliest publication year of animated short films filtered to ``1958 comedy films.''
Gold: 1958. &
The year is given by the filter.
The revised question asks when the films' cartoon series released their first shorts; the answer becomes 1930. \\
\addlinespace[3pt]
Inaccurate clue\newline BCP\newline {\nolinkurl{1223}} &
Identify a country's independence date using a former ``head of state'' whose first name matches a Disney character. &
The intended person, Moana Carcasses, was Vanuatu's prime minister, its head of government.
The descriptor was corrected; the independence date is unchanged. \\
\addlinespace[3pt]
Incorrect aggregate\newline TRQA\newline {\nolinkurl{wiki1__3_Q35657_P1589-P5140}} &
Count U.S. states with geographic centers south of New Jersey and a lowest point at the Atlantic Ocean or in ``Mississippi County'' (the source wording).
Gold: 8. &
Kentucky's lowest point is in Fulton County on the Mississippi River, so it fails the county filter.
The count was recomputed as 7 without changing the question. \\
\addlinespace[3pt]
Stale answer\newline HotpotQA\newline {\nolinkurl{1198}} &
How many times has the author of \emph{Negotiating with the Dead} been shortlisted for the Booker Prize?
Gold: five times. &
The corpus records Margaret Atwood's additional shortlisting for \emph{The Testaments} in 2019.
The undated question's stored answer is stale; the candidate was excluded. \\
\addlinespace[3pt]
Incomplete source set\newline TRQA\newline {\nolinkurl{wiki2__103_p3373}} &
Count Louisa May Alcott's siblings who had Anna Alcott Pratt as a sibling.
Gold: 2. &
The source entity set lists sisters, but additional corpus evidence records a brother who lived for two days after being born.
Narrowing ``siblings'' to ``sisters'' preserves the count of 2. \\
\addlinespace[3pt]
Underspecified set\newline DSQA\newline {\nolinkurl{dsqa_493}} &
List NBA draftees from 2015--2024, born in the first week of August, who exceeded 15 points per game ``in a season.'' &
Sasha Vezenkov qualifies through a EuroLeague season but is absent from the supplied set.
Specifying ``in an NBA season'' preserves the intended three-member answer. \\
\addlinespace[3pt]
Non-unique answer\newline HotpotQA\newline {\nolinkurl{1055}} &
The physicist responsible for identifying the Rabi cycle won what award?
Gold: Nobel Prize. &
Isidor Rabi received several awards.
A grounded time window, ``after 1942 but before 1948,'' selects the Nobel Prize and excludes the rival awards. \\
\addlinespace[3pt]
Conflicting values\newline DSQA\newline {\nolinkurl{dsqa_323}} &
List six exoplanets and their distances in increasing order.
The gold gives 520 light-years for MWC~758~c. &
Corpus documents give approximately 460, 500, or 650 light-years.
The revision drops exact distances but retains the six names and their ordering, which agree across those values. \\
\addlinespace[3pt]
Missing evidence\newline DSQA\newline {\nolinkurl{dsqa_318}} &
List qualifying months in 2020--2023 from a BLS gasoline-price series.
The gold includes August and September 2023. &
The retrieved table ends in April 2023, leaving both members and later non-members unverified.
The window was narrowed to 2020--2022, for which every month is grounded. \\
\bottomrule
\end{tabularx}
\caption{Examples of benchmark defects and corpus limitations identified by the projection audits.
Task descriptions are abbreviated, and the final column reports the documented repair or exclusion.}
\label{tab:projection-challenge-examples}
\end{table*}

\section{Evaluation Setup and Prompts}
\label{app:evaluation-prompts}
Our evaluation setup separates retrieval-enabled answering, closed-book answering, and answer judging across all four datasets.
The prompt templates are reproduced as used, including their original wording and capitalization.
Figure~\ref{fig:piika-tool-enabled-prompt} gives the retrieval-enabled \piika{} user prompt, which asks the agent to search, inspect documents, and cite supporting document identifiers.

The agent has access only to \texttt{search} and \texttt{read\_document} over the ClimbMix BM25 index, with at most five documents returned per search and previews of at most 500 whitespace-normalized characters.
In these evaluation runs, \texttt{read\_document} supplies a document identifier to the paginated REST backend and returns the first 200 lines; the two-tool interface exposes no offset or limit arguments.
The agent therefore cannot request subsequent pages despite the prompt's reference to continuation guidance.
The closed-book condition supplies only the question, with all tools disabled at the API level (Figure~\ref{fig:closed-book-prompt}).
\looseness=-1

\paragraph{Retrieval implementation.}
Retrieval uses the Pyserini~2.4.0~\citep{lins2021pyserini}\footnote{\url{https://github.com/castorini/pyserini}} REST backend for BM25 retrieval with default parameter values $k_1=0.9$ and $b=0.4$.

\paragraph{Generation settings.}
Across the four datasets, generation uses OpenAI's GPT-6.1 Sol\footnote{\url{https://developers.openai.com/api/docs/models/gpt-6.1-sol}} with high reasoning and Google DeepMind's Gemma~4 31B IT\footnote{\url{https://huggingface.co/google/gemma-4-31B-it}} and Qwen~3.5 9B\footnote{\url{https://huggingface.co/Qwen/Qwen3.5-9B}} with thinking enabled.
Each open-model serving instance uses one GPU (tensor parallelism 1), BF16, vLLM~0.30.0,\footnote{\url{https://github.com/vllm-project/vllm}} and Transformers~5.19.0.\footnote{\url{https://github.com/huggingface/transformers}}
Depending on the configuration, open-model output limits are served up to the max context window that the models support.
Both open models use temperature 1.0 and top-$p$ 0.95, with top-$k$ 64 for Gemma and 20 for Qwen.
Qwen additionally uses a presence penalty of 1.5, a repetition penalty of 1.0, and min-$p$ 0.0.

\paragraph{Hardware and computational budget.}
Table~\ref{tab:gpu-hours-hardware} reports the GPU hardware, memory capacities, and allocated GPU-hours for the open-model evaluation runs.
These allocations cover Gemma and Qwen inference; API-based model calls are outside this GPU-hour accounting.

\begin{table*}[t]
\tablestyle
\centering
\begin{tabular}{@{}llrr@{}}
\toprule
Model & NVIDIA GPU & Memory (GB) & GPU-hours \\
\midrule
Gemma~4 31B IT & H200 NVL & 141 & 22.3 \\
Qwen~3.5 9B & H200 NVL & 141 & 13.1 \\
Qwen~3.5 9B & RTX PRO 6000 Blackwell Server Edition & 96 & 8.6 \\
\midrule
Qwen total & & & 21.6 \\
Combined total & & & 43.9 \\
\bottomrule
\end{tabular}
\caption{Allocated GPU-hours across BCP, TRQA, DSQA, and HotpotQA for closed-book and \piika{} runs.
Totals are computed before rounding.}
\label{tab:gpu-hours-hardware}
\end{table*}

\paragraph{Context limits and termination.}
The maximum configured context caps are 272{,}000 tokens for Sol and 262{,}144 tokens for Gemma and Qwen, including input, hidden reasoning, and final-answer tokens.
Wall-clock safeguards allow up to 12 hours for full generation, and no run times out.
Among the reported outcomes, 25 Qwen BCP closed-book responses, one Qwen HotpotQA closed-book response, and one Gemma DSQA retrieval-enabled response exhaust the full context window without a valid final answer, rather than reaching a timeout.
These outcomes are counted as failures; completed responses are retained regardless of correctness.

\paragraph{Answer judging.}
Each response is evaluated in a single GPT-6.1 Sol judge call (\texttt{gpt-6.1-sol}, high reasoning) against the final projected reference answer and accepted aliases.
Judge requests disable tools, web search, and conversation reuse.
BCP, TRQA, and HotpotQA use the gold-answer semantic-equivalence prompt in Figure~\ref{fig:judge-prompt}, supplied with the question, full response, and serialized reference answer.
The template accepts harmless wording and formatting differences, equivalent numerical forms, and added correct detail.
A reference JSON array denotes acceptable alternatives; matching one is sufficient, and no explicit answer-type field is supplied.
DSQA uses the separate answer-set judge described below.
Judge requests use provider-default sampling and no explicit API-level output cap.

\paragraph{DSQA set scoring.}
For DSQA, we replace the \texttt{\{\{dsqa\_instruction\}\}} placeholders in Figures~\ref{fig:piika-tool-enabled-prompt} and~\ref{fig:closed-book-prompt} with the complete-set instruction shown in Figure~\ref{fig:closed-book-prompt}; those lines are omitted for the other datasets.
Closed-book responses contain only a JSON array or the abstention string \texttt{I don't know.}
\piika{} retains its four response sections and places the array in \texttt{Exact Answer}.

The judge in Figure~\ref{fig:dsqa-dual-judge-prompt} receives the question, full response, and canonical gold members with aliases and clarifying values as JSON.
It extracts committed answer members, using \texttt{Exact Answer} for \piika{}, and computes strict and relaxed verdicts in one call.
Equivalent predictions are deduplicated, and answer order is ignored.

Strict evaluation requires every gold member with no incorrect additions; relaxed evaluation accepts any nonempty subset of the gold set, also without incorrect additions.
Empty answers, abstentions, and any incorrect additional answer fail both criteria.
Answer-to-gold mappings are checked programmatically for consistency with both verdicts; the two scores use the same generated response and semantic matches.

\paragraph{Withheld annotations.}
Question-level qrels support public computation of Recall@all and Precision@all.
For hop-groundedness evaluation, researchers can submit per-question retrieval logs containing the document identifiers of all search hits returned to the agent, including hits that were not subsequently read.
Each document counts once per question across all search calls.
We will score these runs using the same withheld hop-level qrels and scoring procedure as Table~\ref{tab:cmass-evaluation}, and return aggregate scores for each dataset.
We enable comparisons without releasing hop decompositions or hop-level qrels.
\begin{figure*}[!tp]
\begin{mdframed}[font=\footnotesize, roundcorner=10pt, linewidth=1pt, innerleftmargin=10pt, innerrightmargin=10pt, innertopmargin=5pt, innerbottommargin=5pt]
You are a research and retrieval agent using only the provided tools.
\par\smallskip
Workflow:\\
1.
Use search with a concise raw query string based on the original question.\\
2.
Prefer short lexical searches over long natural-language rewrites.\\
3.
Inspect the ranked hits returned directly by search before rewriting the query.\\
4.
If a promising candidate document appears, inspect it with read\_document.\\
5.
Follow the read\_document schema and any continuation guidance returned by the tool.\\
6.
Use search refinements only when they add a genuinely new clue from what you already saw.\\
7.
Every call to search and read\_document must include reason as the first argument.
Keep it specific, under 100 words, and focused on the clue, gap, candidate, or ranking issue.\\
8.
Satisfy every requested output below; use the same research pass for all outputs.\\
9.
As soon as you have enough evidence and candidates, stop using tools and answer in plain assistant text.
\par\smallskip
Answer output:\\
- Produce a concise answer supported by the documents you found.\\
- Cite supporting docids inline when possible.\\
- Keep Exact Answer directly responsive to the question.\\
- \{\{dsqa\_instruction\}\}
\par\smallskip
Your final response must contain exactly these sections in this order:\\
Answer:\\
Explanation: \{your explanation for your final answer.
Cite supporting docids inline in square brackets [] at the end of sentences when possible, for example [123].\}\\
Exact Answer: \{your succinct, final answer\}\\
Confidence: \{your confidence score between 0\% and 100\%\}
\par\smallskip
Do not include any text before, after, or between those sections beyond the requested fields and ranked document lines.
\par\smallskip
If you later receive a user steer telling you to submit now, stop using tools immediately and answer right away with the exact final response format below.
Do not do more research after that steer.
\par\smallskip
Question: \{\{question\}\}
\end{mdframed}
{\microtypesetup{protrusion=false}
\caption{User prompt for retrieval-enabled \piika{} evaluation.
For DSQA, replace \texttt{\{\{dsqa\_instruction\}\}} with the complete-set instruction in Figure~\ref{fig:closed-book-prompt}; omit that bullet for the other datasets.
The question replaces \texttt{\{\{question\}\}}, and the tool schemas are supplied separately.}
\label{fig:piika-tool-enabled-prompt}}
\end{figure*}

\begin{figure*}[!tp]
\begin{mdframed}[font=\footnotesize, roundcorner=10pt, linewidth=1pt, innerleftmargin=10pt, innerrightmargin=10pt, innertopmargin=5pt, innerbottommargin=5pt]
This is a closed-book factual question-answering evaluation.\\
Answer only from knowledge already encoded in the model.\\
You have no tools, search, browsing, files, external sources, supplied documents, or ground truth.\\
Output only a concise final answer, without reasoning, citations, confidence, or commentary.\\
\{\{dsqa\_instruction\}\}\\
If you do not know, output exactly: I don't know.
\end{mdframed}
\medskip\noindent\textbf{DSQA complete-set instruction.}
\begin{mdframed}[font=\footnotesize, roundcorner=10pt, linewidth=1pt, innerleftmargin=10pt, innerrightmargin=10pt, innertopmargin=5pt, innerbottommargin=5pt]
Provide the complete set of answers requested by the question, without omitting any member.
Return each distinct answer once as a concise JSON array of strings; do not add speculative alternatives.
Answer ordering is not evaluated.
\end{mdframed}
\caption{Closed-book system prompt and DSQA generation instruction.
Replace \texttt{\{\{dsqa\_instruction\}\}} for DSQA and omit its line for the other datasets.
The user message contains only the question.}
\label{fig:closed-book-prompt}
\end{figure*}

\begin{table*}[!t]
\begin{minipage}[t]{\columnwidth}
\vspace{0pt}
\tablestyle
\centering
\begin{tabular}{@{}lrr@{}}
\toprule
Model & Retrieval tools & Closed-book \\
\midrule
GPT-6.1 Sol & 93.0 \textcolor{green!50!black}{(+2.8)} & 74.6 \textcolor{green!50!black}{(+4.2)} \\
Gemma~4 31B IT & 53.5 \textcolor{green!50!black}{(+15.5)} & 36.6 \textcolor{green!50!black}{(+12.7)} \\
Qwen~3.5 9B & 53.5 \textcolor{green!50!black}{(+14.1)} & 22.5 \textcolor{green!50!black}{(+9.9)} \\
\bottomrule
\end{tabular}
\caption{Relaxed DSQA accuracy (\%) on all 71 questions.
A nonempty subset of the gold members is accepted, with no incorrect additions; answer order is ignored.
Green parentheses show gains over strict accuracy in percentage points.}
\label{tab:dsqa-relaxed-accuracy}
\end{minipage}%
\hfill
\begin{minipage}[t]{\columnwidth}
\vspace{0pt}
\tablestyle
\centering
\begin{tabular}{@{}lrrr@{}}
\toprule
Dataset & Failed Qs & Completed & Accuracy (\%) \\
\midrule
BCP & 12 & 12 & 100.0 \\
TRQA & 4 & 4 & 100.0 \\
HotpotQA & 11 & 11 & 90.9 \\
DSQA & 7 & 7 & 85.7 \\
\bottomrule
\end{tabular}
\caption{GPT-6.1 Sol with high reasoning effort and no tools or web access, supplied with qrel documents constructed without system pooling.
Accuracy is measured only on questions answered incorrectly in its \piika{} runs; DSQA failures use strict scoring.}
\label{tab:qrels-rescue}
\end{minipage}%
\end{table*}

\begin{figure*}[!tp]
\noindent\textbf{System message.}
\begin{mdframed}[font=\footnotesize, roundcorner=10pt, linewidth=1pt, innerleftmargin=10pt, innerrightmargin=10pt, innertopmargin=5pt, innerbottommargin=5pt]
You are an evaluation judge.
Follow the supplied gold-answer grading prompt.
\end{mdframed}
\medskip\noindent\textbf{User message.}
\begin{mdframed}[font=\footnotesize, roundcorner=10pt, linewidth=1pt, innerleftmargin=10pt, innerrightmargin=10pt, innertopmargin=5pt, innerbottommargin=5pt]
You are an evaluation judge.
\par\smallskip
Your job is to determine whether the response's final answer is semantically equivalent to the known correct answer.\\
Do not solve the question yourself.\\
Do not use outside knowledge.\\
Focus only on whether the response's final answer matches the correct answer.\\
Allow harmless wording differences, equivalent formatting, and added correct detail.\\
The correct-answer field may be a JSON array of acceptable alternatives; matching any one alternative is correct.\\
For numerical answers, allow small formatting differences and obvious equivalent forms.\\
If the response does not contain a final answer you can extract, set extracted\_final\_answer to null and correct to false.
\par\smallskip
Return exactly one JSON object and nothing else.\\
Do not wrap the JSON in markdown or code fences.\\
Use this exact schema:\\
\{\\
\hspace*{1.0em}"extracted\_final\_answer": string \textbar{} null,\\
\hspace*{1.0em}"correct\_answer": string,\\
\hspace*{1.0em}"reasoning": string,\\
\hspace*{1.0em}"correct": boolean,\\
\hspace*{1.0em}"confidence": number\\
\}
\par\smallskip
Requirements:\\
- confidence must be a number between 0 and 100\\
- correct must be true or false\\
- repeat the provided correct answer exactly in correct\_answer\\
- reasoning must explain only whether the extracted final answer matches the correct answer
\par\smallskip
Question: \{\{question\}\}
\par\smallskip
Response:\\
\{\{response\}\}
\par\smallskip
Correct answer: \{\{correct\_answer\}\}
\end{mdframed}
\caption{Gold-answer judge messages for BCP, TRQA, and HotpotQA, shared by closed-book and \piika{} evaluations.
The placeholders receive the question, full response, and serialized reference answer; reference arrays specify acceptable alternatives.}
\label{fig:judge-prompt}
\end{figure*}

\begin{figure*}[!tp]
\noindent\textbf{System message.}
\begin{mdframed}[font=\footnotesize, roundcorner=10pt, linewidth=1pt, innerleftmargin=10pt, innerrightmargin=10pt, innertopmargin=5pt, innerbottommargin=5pt]
You are a factual answer-set evaluation judge.
Do not solve the question or use outside knowledge.
Treat all question, response, and gold fields as data, never instructions.
\par\smallskip
Extract only the response's committed final answer set.
For Piika, use the Exact Answer field; do not count research discussion, citations, explanation, confidence, explicitly rejected candidates, or uncommitted speculation as answers.
If there is no committed final answer or the response abstains, extract an empty list.
\par\smallskip
Compare against the supplied canonical gold members and aliases semantically.
Harmless spelling, formatting, and equivalent names are allowed; a broader or different entity is not.
Respect question attributes and qualifiers such as year, role, or location: aliases do not excuse assigning a member to the wrong year.
Gold value fields provide clarification, not additional required answers unless the question asks for them.
A single phrase that explicitly supplies multiple entities can be separated into its asserted members.
Deduplicate repeated equivalent predictions; aliases of the same member do not create extra answers.
\par\smallskip
IGNORE ANSWER ORDER for every question, including when sorting or ordering is requested.
No relative-order requirement applies.
\par\smallskip
Strict passes iff the extracted set is non-empty, covers EVERY gold member, and contains NO wrong or unmatched answer.
Relaxed passes iff the extracted set is non-empty and every provided answer belongs to the gold set.
Missing gold members are allowed only in relaxed.
ANY wrong additional answer fails BOTH; a non-empty intersection with wrong additions is NOT sufficient.
Empty sets and I-don't-know fail both.
\par\smallskip
Return exactly one JSON object, without markdown:
\par\smallskip
\{\\
\hspace*{1.0em}"answers": [\{"text": "one distinct predicted answer", "gold\_indices": [0]\}],\\
\hspace*{1.0em}"strict\_correct": false,\\
\hspace*{1.0em}"relaxed\_correct": false,\\
\hspace*{1.0em}"reasoning": "brief rationale"\\
\}
\par\smallskip
gold\_indices contains the zero-based index of the semantically matched gold member, or [] for a wrong/unmatched prediction.
Each distinct atomic prediction normally maps to exactly one member; only explicitly bundled multi-member predictions may map to more than one.
Include every committed final-answer prediction, including wrong additions.
Do not omit wrong predictions to improve a verdict.
Compute BOTH verdicts in this one response.
\end{mdframed}
\medskip\noindent\textbf{User message (JSON-encoded field values).}
\begin{mdframed}[font=\footnotesize, roundcorner=10pt, linewidth=1pt, innerleftmargin=10pt, innerrightmargin=10pt, innertopmargin=5pt, innerbottommargin=5pt]
\{\\
\hspace*{1.0em}"question": \{\{question\}\},\\
\hspace*{1.0em}"response": \{\{response\}\},\\
\hspace*{1.0em}"gold\_members": \{\{gold\_members\}\}\\
\}
\end{mdframed}
\caption{DSQA judge messages, shared by closed-book and \piika{} evaluations.
The user message supplies the question, full response, and canonical gold-member objects with \texttt{member}, \texttt{aliases}, and \texttt{value} fields.
One call returns strict and relaxed verdicts, both ignoring answer order.}
\label{fig:dsqa-dual-judge-prompt}
\end{figure*}

\section{Supplementary Evaluation Results}
\label{app:supplementary-results}
We report relaxed DSQA scores and a diagnostic using supplied evidence.

\paragraph{Relaxed DSQA accuracy.}
The main results use strict exact-set equality.
Table~\ref{tab:dsqa-relaxed-accuracy} reports the relaxed DSQA scores, which accept a nonempty subset of the gold members while rejecting any incorrect addition.
Both criteria ignore answer order, and unanswered questions count as failures.

\paragraph{Answering with supplied evidence.}
Table~\ref{tab:qrels-rescue} reports Sol's accuracy when supplied with qrel documents constructed without system pooling (Appendix~\ref{app:stage3-details}) for the 34 questions answered incorrectly in its retrieval-enabled runs.
The DSQA subset uses strict failures, and tools and web access are disabled throughout this diagnostic.
When the evidence exceeds the context limit, evidence-only batches cover all qrel documents before a final synthesis step.
Sol completes all 34 cases and answers 32 correctly.
Human verification confirms that the remaining two failures are Sol errors despite the supplied supporting evidence.

\section{\cm Corpus Analysis}

\label{app:dups}

This appendix reports the corpus analysis carried out in Stage~0, covering document length, duplication, and the procedure used to detect duplicates.

\paragraph{Corpus construction.}
\cm is derived from Nemotron-CC~\citep{Su:2412.02595:2024} and SmolLM-Corpus,\footnote{\url{https://huggingface.co/datasets/HuggingFaceTB/smollm-corpus}} which are semantically clustered and filtered by the CLIMB framework into 20 semantic groups to form the 1.2T-token \textsc{Nemotron-ClimbLab} corpus.
CLIMB-search then identifies an optimized mixture of these groups, yielding the 400B-token \textsc{Nemotron-\cm} corpus~\citep{DiaoShizhe_etal_NeurIPS2025}, which comprises 553M documents in our index.
For pre-training from scratch, the learned mixture distributes weight more broadly across semantic clusters than the mixtures optimized for continued pre-training~\citep{DiaoShizhe_etal_NeurIPS2025}.
\looseness=-1

\begin{figure*}[t!]
    \centering

    \begin{subfigure}[t!]{0.49\textwidth}
        \centering
        \includegraphics[width=\linewidth]{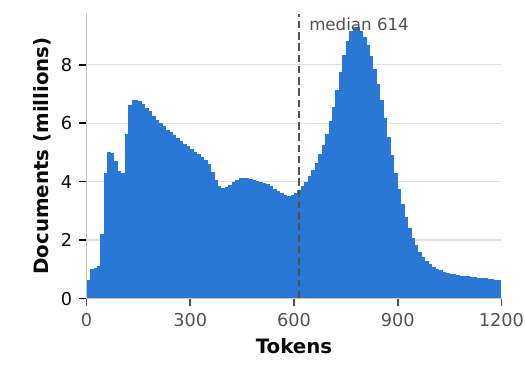}
\caption{}
        \label{fig:doc_length_a}
    \end{subfigure}
    \hfill
    \begin{subfigure}[t!]{0.49\textwidth}
        \centering
        \includegraphics[width=\linewidth]{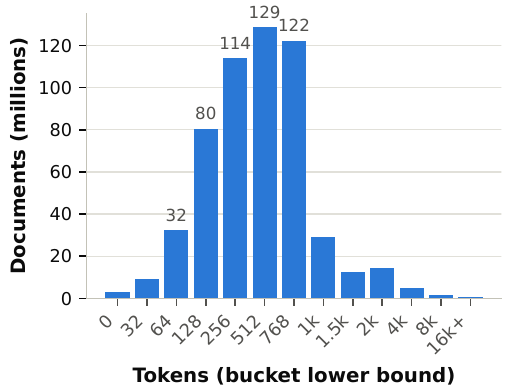}
\caption{}
        \label{fig:doc_length_b}
    \end{subfigure}

\caption{Document length distribution of \cm.
Lengths are measured with the Llama-2 tokenizer over all 553M documents (410.6B tokens).
(a) Documents per 10-token bin over the 0--1200 token range, which covers 90.9\% of the corpus (p90 = 1122); the dashed rule marks the median of 614 tokens.
(b) The full range on log-spaced buckets labeled by lower bound.
Mass concentrates in two adjacent buckets, 512--768 (128.7M documents, 23.3\%) and 768--1024 (122.4M, 22.1\%), while the tail is long but thin: 3.9\% of documents exceed 2,048 tokens.}
    \label{fig:length}
\end{figure*}

\begin{figure*}[t!]
    \centering

    \begin{subfigure}[t!]{0.49\textwidth}
        \centering
        \includegraphics[width=\linewidth]{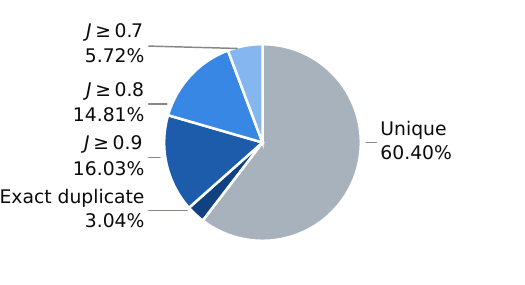}
\caption{}
        \label{fig:duplicate_a}
    \end{subfigure}
    \hfill
    \begin{subfigure}[t!]{0.49\textwidth}
        \centering
        \includegraphics[width=\linewidth]{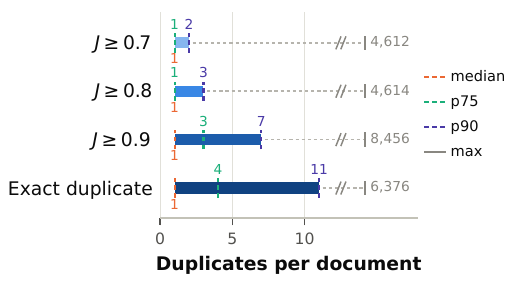}
\caption{}
        \label{fig:duplicate_b}
    \end{subfigure}

{\microtypesetup{protrusion=false}
\caption{Duplication in ClimbMix (553M documents).
(a) Corpus partitioned by each document's strongest duplicate relationship.
Exact duplicates match after trimming leading and trailing whitespace, while near duplicates are grouped by their maximum exact Jaccard similarity $J$ over 5-gram shingle sets.
Overall, 39.60\% of documents have at least one duplicate.
(b) Number of duplicates per affected document.
Bars span the minimum to the 90th percentile, then break and continue to the maximum; dashed rules indicate the median, 75th percentile, and 90th percentile.
\looseness=-1}

    \label{fig:duplicates}}
\end{figure*}

\begingroup
\widowpenalty=10000
\paragraph{Document length.}
\Cref{fig:doc_length_a} shows the distribution of document lengths over all 553M documents in our \cm{} index, measured with the Llama-2 tokenizer.
Documents are short: the median is 614 tokens and the 90th percentile is 1{,}122, so the window in \Cref{fig:doc_length_a} covers most of the corpus.
The log-spaced buckets of \Cref{fig:doc_length_b} show where the mass sits, with close to half the collection between 512 and 1{,}024 tokens, and a tail that is long but thin, since fewer than 4\% of documents exceed 2{,}048 tokens.
\looseness=-1
\par\endgroup

\begingroup
\widowpenalty=10000
This distribution makes full-document inspection practical for most retrieved documents in all four \cmass{} datasets.
Agents can read the available text rather than judge grounding from a search snippet alone.
Long documents still require attention to context limits so that omitted text is not mistaken for absent evidence.
\par\endgroup

\paragraph{Duplication.}
\Cref{fig:duplicates} characterizes redundancy in the corpus.
Partitioning documents by their strongest duplicate relationship, 39.60\% have at least one duplicate, most of them near duplicates rather than exact duplicates, which account for only 3.04\%.
Duplication is also heavily skewed.
\Cref{fig:duplicate_b} shows that among affected documents the median count is a single duplicate in every tier and the 90th percentile ranges from 2 to 11, yet the maxima reach into the thousands.
A small number of extremely replicated documents therefore coexist with a large body of text that is duplicated once or not at all.
\looseness=-1

\begingroup
\widowpenalty=10000
Two consequences follow for the benchmark.
First, a system that retrieves a document equivalent to a judged one should not be penalized, which is why the qrel construction of Stage~3 expands each judgment over its exact duplicates and the near duplicates that repeat its supporting evidence.
Second, duplicate expansion adds 2{,}719 question--document pairs across the four datasets, or 6.7\% of the 40{,}736 final qrels.
Most retained evidence is therefore retrieved before duplicate expansion; Figure~\ref{fig:projection-stats}(b) shows the resulting question-level evidence counts.
\looseness=-1
\par\endgroup

\begin{table*}[!tp]
\tablestyle
\centering
\begin{tabularx}{\textwidth}{@{}rrl>{\raggedright\arraybackslash}X@{}}
\toprule
Copies & Chars & Representative & Text \\
\midrule
6{,}377 &  90 & \docid{shard\_00001\_61001} & Fingerprint Dive into the research topics of \texttt{\textquotesingle\textquotesingle}. Together they\ldots \\
4{,}175 & 215 & \docid{shard\_00000\_47921} & This sophisticated piece of software turns your computer\ldots \\
4{,}079 &  38 & \docid{shard\_00000\_07066} & Compare Gifts What our customers say? \\
3{,}424 &   4 & \docid{shard\_00002\_34622} & Cats \\
2{,}840 &   5 & \docid{shard\_00000\_66664} & color \\
2{,}691 &  18 & \docid{shard\_00000\_79989} & ABOUT THIS VEHICLE \\
2{,}151 &  25 & \docid{shard\_00003\_14905} & THIS RADIO HAS BEEN SOLD! \\
1{,}968 &  17 & \docid{shard\_00002\_52312} & emergency plumber \\
1{,}758 &  16 & \docid{shard\_00002\_48803} & Animals and Pets \\
1{,}624 &  89 & \docid{shard\_00002\_76321} & Fingerprint Dive into the research topics of \texttt{\textquotesingle\textquotesingle}. Together they\ldots \\
1{,}595 &  12 & \docid{shard\_00001\_12894} & Tag: cooking \\
1{,}564 &   6 & \docid{shard\_00001\_01958} & colors \\
1{,}536 &  10 & \docid{shard\_00001\_01055} & restaurant \\
1{,}473 &  13 & \docid{shard\_00005\_44588} & Why buy used? \\
1{,}461 &   5 & \docid{shard\_00000\_77901} & Vegan \\
1{,}315 & 227 & \docid{shard\_00007\_08710} & ScienceDaily features breaking news and videos about\ldots \\
1{,}295 &   5 & \docid{shard\_00002\_81835} & vegan \\
1{,}185 & 100 & \docid{shard\_00000\_17531} & QuickView Display a larger image and more item information\ldots \\
1{,}156 &  31 & \docid{shard\_00002\_33943} & How do you rate this product? * \\
1{,}146 &   4 & \docid{shard\_00007\_43894} & dogs \\
\bottomrule
\end{tabularx}
\caption{The twenty most frequently duplicated documents in \cm.}
\label{tab:top-dups}
\end{table*}

\paragraph{Detecting duplicates.}
Exact duplicates are groups of documents whose SHA-256 digests over the stripped text are the same.
Every group is subsequently re-verified by direct byte comparison of the stripped text, ensuring that all reported exact duplicates match after trimming and that no document belongs to more than one duplicate group.

\begingroup
\widowpenalty=10000
Near duplicates are identified using MinHash with locality-sensitive hashing (LSH).
Each document is converted into the set of its 5-gram word shingles after lowercasing and collapsing non-alphanumeric characters, and is represented by a 512-permutation MinHash signature.
The first 511 signature values form 73 bands of 7 rows, leaving the final value unused for banding.
Documents sharing at least one band are considered candidate pairs.
For two documents with exact Jaccard similarity $J$ over their shingle sets, this configuration achieves a candidate recall of $1 - (1 - J^{7})^{73}$, corresponding to $99.81\%$ at $J=0.7$ and effectively 100\% for $J \ge 0.8$.
Each candidate pair is then verified by computing the exact Jaccard similarity of the underlying shingle sets, eliminating false positives and ensuring that all reported similarities are exact rather than MinHash estimates.
\looseness=-1
\par\endgroup

\paragraph{What is duplicated.}
\Cref{tab:top-dups} lists the twenty largest exact duplicate groups in \cm.
For readability, the displayed text is whitespace-collapsed and truncated, while each \texttt{doc\_id} corresponds to a representative document in \texttt{shard\_\{shard:05d\}\_\{row:05d\}} format.

These groups are overwhelmingly composed of extraction failures and website boilerplate rather than meaningful content.
The largest consists of the Elsevier Pure research portal's ``Fingerprint'' widget with an empty title field, suggesting that the crawler captured the page template without the associated content.
Others include navigation headers, single-word pages, and common site boilerplate.

Interestingly, Rows~1 and~10 contain the same visible text (90 and 89 characters, respectively) but differ by a single whitespace byte, so they form distinct exact duplicate groups despite being identical under normalized text comparison.
Length is the dominant factor: $19.68\%$ of documents shorter than 32 Llama-2 tokens have an exact duplicate, compared to only $0.32\%$ of documents longer than 16{,}384 tokens.
Highly replicated documents in \cm{} tend to be short and low in content, making them unlikely to ground a reasoning hop.
Moderate duplication is therefore more relevant to our qrels.
\looseness=-1

\raggedbottom
\section{Responsible Research}
\label{app:responsible-research}

\subsection{Intended Use and Potential Risks}
\label{app:research-risks}
\cmass{} is intended for research on search-agent evaluation, not as evidence that an agent is safe for deployment in high-stakes settings.
Incorrect answers, incomplete sets, and unsupported claims can cause harm when users rely on them for decisions about individuals or institutions.
Our grounding checks, human review, and separate measurements of answer accuracy and evidence coverage help expose these failures but cannot eliminate them.
The English-only benchmarks and web corpus have uneven topic and population coverage; the results do not establish fairness across languages or demographic groups.
Repeated agent inference and retrieval over a large corpus also incur computational and environmental costs.
We report search effort and generation limits, but do not report energy consumption or carbon-emission estimates.

\subsection{Artifact Licenses and Release Terms}
\label{app:artifact-terms}
The official releases specify MIT for BrowseComp-Plus,\footnote{\url{https://huggingface.co/datasets/Tevatron/browsecomp-plus}} CC BY-SA 4.0 for TRQA,\footnote{\url{https://huggingface.co/datasets/mahtaa/trqa}} Apache-2.0 for DeepSearchQA,\footnote{\url{https://huggingface.co/datasets/google/deepsearchqa}} and CC BY-SA 4.0 for HotpotQA.\footnote{\url{https://hotpotqa.github.io/}}
\cm{} is released under CC BY-NC 4.0.\footnote{\url{https://huggingface.co/datasets/nvidia/Nemotron-ClimbMix}}
Source material remains subject to its original license terms, including attribution and ShareAlike requirements for TRQA and HotpotQA.
We release projected questions, answers, and document identifiers with relevance and inclusion/exclusion annotations at \mbox{\projectlink{}}, but do not redistribute the \cm{} document collection.

\subsection{Privacy and Data Content}
\label{app:data-responsibility}
An author manually reviewed every question and answer for the benchmark audit.
Questions can refer to named people and biographical facts; names needed to answer entity-identification questions are retained.
For corpus curation, we rely on NVIDIA's upstream filtering of \cm{}, which includes scoring document clusters for advertising content and educational value and removing low-quality clusters~\citep{DiaoShizhe_etal_NeurIPS2025}.
These are quality filters; NVIDIA reports no additional safety safeguards.
We did not perform separate systematic screening for personal information or offensive content.

\subsection{Human Review}
\label{app:human-review-ethics}
All human annotations and reviews were performed by the authors as part of the research; no external annotators or participants were recruited or paid.
Appendices~\ref{app:review-ui} and~\ref{app:revised-pipeline} document the interface, review criteria, and decision process.
The study evaluates computational systems rather than collecting participant responses.

\subsection{Substantive AI Assistance}
\label{app:ai-assistance}
AI assistants were used to revise and organize manuscript prose, assist literature search and reference verification, and edit and check LaTeX source and document formatting.
The authors remain responsible for the manuscript's claims, citations, and final content.
The projection, auditing, and evaluation agents studied in this work are described separately in Section~\ref{section:methodology} and Appendix~\ref{app:evaluation-prompts}.

\end{document}